\documentclass[apj]{emulateapj}

\def\RSUN{R$_{\sun}$ }
\def\DEG{$^\circ$} 
\def\LL{\mbox{$\:\lambda\lambda $ }}
\def\kms{$\rm km~s^{-1}$ }
\def\L{$\lambda$ }
\def\LA{Ly$\alpha$ }
\def\LB{Ly$\beta$ }

\def\C3{$\rm C~III$ }
\def\fe{$\rm [Fe~XVIII]$ }
\def\ca{$\rm [Ca~XIV]$ }
\def\Osix{$\rm O~VI$ }
\def\Ofiv{$\rm O~V$ }
\def\Si12{$\rm Si~XII$ }
\def\inte{$\rm ph~s^{-1}cm^{-2}sr^{-1}$ }
\def\flux{$\rm ph~cm^{-2}s^{-1}$ }

\begin{document}

\title{Bright Ray-like Features in the Aftermath of CMEs: White Light vs UV Spectra}

\author{A. Ciaravella\altaffilmark{1}, 
D. F. Webb\altaffilmark{2}, 
S. Giordano\altaffilmark{3} 
\& J.C. Raymond\altaffilmark{4}
}

\altaffiltext{1}{INAF-Osservatorio Astronomico di Palermo, P.za Parlamento 1, 90134 Palermo, Italy}
\altaffiltext{2}{Institute for Scientific Research, Boston College, Newton, MA 02459, USA}
\altaffiltext{3}{INAF-Osservatorio Astrofisico di Torino, via Osservatorio 20, 10025 Pino Torinese, Italy}
\altaffiltext{4}{Harvard-Smithsonian Center for Astrophysics, 60 Garden Street, Cambridge, MA 02138, USA}

\begin{abstract}

Current sheets are important signatures of magnetic reconnection in the
eruption of confined solar magnetic structures. Models of Coronal Mass Ejections (CMEs) involve formation 
of a current sheet connecting the ejected flux rope with the post eruption magnetic loops. 
Current sheets have been identified in white light images of CMEs as narrow rays
trailing the outward moving CME core, and in ultraviolet spectra as 
narrow bright features emitting the \fe line.
In this work samples of rays detected in white light images or in ultraviolet spectra have been
analyzed.  Temperatures, widths, and line intensities of the rays have been measured, and 
their correlation to the CME properties has been studied. 
The samples show a wide range of temperatures with hot, coronal and cool rays.  In some cases, the
UV spectra support the identification of rays as current sheets, but they show that some
white light rays are cool material from the CME core. In many cases,
both hot and cool material are present, but offset from each other along the UltraViolet Coronagraph Spectrometer 
(UVCS) slit.
We find that about 18\% of the white light rays show very hot gas
consistent with the current sheet interpretation, while about 23\% show cold gas that we attribute
to cool prominence material draining back from the CME core.  The remaining events have ordinary
coronal temperatures, perhaps because they have relaxed back to a quiescent state.
 
\end{abstract}
\keywords{Sun: coronal mass ejections (CMEs), UV radiation, Corona}

\section{Introduction}\label{intro}

The loss of equilibrium in magnetic structures that disrupts coronal structures and 
ejects confined plasma into interplanetary space requires magnetic reconnection 
\citep{For00,For06,Lo01}. 
Dissipation of magnetic energy and topological reconfiguration of the magnetic field 
arise in strong current sheets, regions of large gradient in the tangential component 
caused by layers of antiparallel field lines.
In the standard picture of solar eruptions, a current sheet connecting the flare loops
to the CME core is where magnetic free energy is converted into thermal and bulk kinetic
energies and into beams of energetic particles \citep{Lit96,Lin00,For06}. The reconnection 
must be rapid in order to explain the rapid rise times of flares, and current models stress
kinetic effects \citep{Pr00}, turbulence \citep{laz99,Em04} or instabilities related to the
tearing mode and the growth of magnetic islands \citep{lou07, ji11}.  To test these theories
it is important to measure the physical parameters, such as temperatures, densities, 
velocities, turbulence and widths, of the current sheets.

Recent observational studies of current sheets include coronagraph observations 
in white light based on timing and morphology of rays that connect the eruption
site to the apparent trailing edge of a CME flux rope \citep{W95,W03} or
based on coincidence of a bright feature with a current sheet identified from
X-rays \citep{pat11}.  \cite{vrs09} measured the properties
of several such rays as functions of height and time.  \cite{song12} analyzed the morphology of 
11 white light rays that had plasma blobs flowing outwards sequentially along the ray. 
Four of their rays are also included in the sample studied in this paper.

In ultraviolet spectra, current sheets
were identified as narrow regions of very hot plasma seen with UVCS between flare 
loops and CME cores \citep{Cia02, Ko03, Be06, Cia08, Sch10}
or as hot gas with very large line widths above flare loops observed by
the Solar Ultraviolet Measurements of Emitted Radiation (SUMER), \citep{in03, Wa07}.  
Narrow band EUV images also show 
hot gas in current sheets above the limb, notably the Transition Region and Coronal Explorer images
of \cite{in03}  and the Solar Dynamics Observatory/Atmospheric Imaging Assembly (AIA) images of \cite{re11} and \cite{cheng}.
Current sheets have also been identified with X-ray emission above the flare
loops \citep{sui03, sav10, sav11}.
 
In this paper we attempt to clarify the relationship between the white light
and UV manifestations of current sheets, both to determine whether the identified
features really are current sheets and to better constrain their physical
properties.  We approached the problem from two directions.  First, we
systematically searched for white light rays that meet criteria
for identification as current sheets: The requirements of a 'disconnection event'
morphology and location in wake of a CME led to 157 candidate current sheets
from among many thousands of radial features in LASCO images. 
We then examined UVCS data to see whether
they contained hot gas.  Second, we used the catalog of UVCS observations of
CMEs in which current sheets or possible current sheets were indicated by
narrow emission features in the [Fe XVIII] line.  Again, this was a stringent
criterion, as fewer than 20 such events have been found among the thousand
CMEs in the UVCS CME catalog. We then
examined the LASCO data for these events to see whether they corresponded to white light
rays that met the current sheet criteria.  The white light selected rays
cover the periods 1996 April through 1998 December to sample solar
minimum and the year 2001 to sample solar maximum.  The UV selected rays
were obtained from the period 1996 June through 2005 December. 

The paper is organized as follows. In Section~\ref{uvlasco} 
the UV and White Light (WL) instruments used for the analysis are described. 
WL and UV selection of the rays is presented in Section~\ref{raysel}. The analysis
and results are in Section~\ref{anal}, and summary and conclusions are in Section~\ref{concl}.

\section{UVCS and LASCO Observations}\label{uvlasco}

The study we present in this paper is mainly based on the observations of two instruments 
aboard the Solar and Heliospheric Observatory, the Large Angle and Spectrometric Coronagraph
Experiment (LASCO) and UVCS.
LASCO C2 is a white light coronagraph, that continuously observes the solar corona 
from 2.3 to 6 \RSUN with a cadence of about 20 min and a pixel size of 11.4$^\prime$$^\prime$ \citep{Br95}.
For those events in which the UVCS slit was located below the LASCO C2 occulter disk we also examined, whenever 
the data were available, the Mark-III or Mark-IV K-coronameter data from Mauna 
Loa Solar Observatory (MLSO). The MLSO coronagraphs are ground-based instruments that produce polarization 
brightness maps of the solar corona from 1.14 to 2.86 \RSUN in the range 7000 - 9000 \AA\/ 
with a 3 min cadence. The typical observing period is from $\sim$ 17:00 to $\sim$ 02:00 UT each day.  The data for some days show
misalignment with LASCO images as large as 13\DEG.  In those cases, we adjusted the MLSO images by eye.

Unlike the WL coronagraph with its complete view of the solar corona, the UVCS spectrometer has 
a field of view defined by its narrow entrance slit 42$^\prime$ long and up to 82$^\prime$$^\prime$ wide. 
The slit can be placed at any polar angle and at heliocentric heights between 1.5 and 9 \RSUN. UVCS has two UV channels optimized to 
detect the \Osix \LL1032, 1037 doublet and HI \LA lines, respectively, but it covers  the 
wavelength range from 945  to 1270 \AA\/ in first order and  from 473 to 635 \AA\/ in second order
(see \citealt{K95} for a detailed description). The OVI channel has a redundant path that allows
detection of the HI \LA line.  The spatial pixels are 7$^\prime$$^\prime$ and spectral pixels are 0.0993 \AA\ (0.0915 \AA\ 
for the redundant path) and 0.1437\AA\/ for the OVI and LYA channels, respectively.  The radiometric calibration 
is discussed by \cite{G02}.  The second order calibration needed for the Si XII lines
is uncertain, but we adopt factors of 0.14 and 0.09 lower efficiency for the \LL521 and 499 lines, respectively.  Due
to telemetry limitations, UVCS can only observe the full wavelength range using low spatial and spectral resolution. 
A full range is usually obtained with a 10 by 2 pixel binning in the spatial and spectral ranges respectively.  
However, most of the UVCS observations are made with spatial resolution of 3 (21$^\prime$$^\prime$) or 6 
(42$^\prime$$^\prime$) pixels and a reduced spectral range. Observations with high spectral and spatial 
resolution require masking the 
detector in order to select the wavelength range of interest with the desired spectral and spatial binning.

The UVCS daily observation schedule includes a synoptic program in which the corona is scanned between 1.5 and 
3 \RSUN at eight polar angles (PA) in about 10 hours, and special observations designed to observe specific 
objectives, such as streamers, coronal holes, CMEs etc.
The synoptic program has changed over the years, with varying ranges of heliocentric 
heights observed, spatial and spectral resolutions 
and spectral range. Despite this, the spectra always included the  O VI doublet, Si XII \LL 499 or 521, \LA, \LB and C III \L977.  
During most of
2001, the synoptic scans also included the high temperature line of \fe (10$^{6.8}$ K), 
but in 1996 through 1998 they did not. \fe is the key indicator of hot gas because it is 
only detected by UVCS above 1.5 \RSUN in the wake of CMEs as 
bright spatially narrow features \citep{Cia02, Ko03, Be06, Cia08, Sch10} or hot blobs \citep{ray03}.

Streamer studies, especially those devoted to elemental abundances, have a large wavelength coverage, often the full wavelength range.
CME watch sequences are designed to have high cadence with typical exposure time of 120 or 200 s. The observations are at a constant
height or switch between two heights. Typical observation heights are between 1.6 and 2.3 \RSUN, and they cover a broad ionization range from 
cool lines such as C III ( $\sim$ 10$^5$ K) to hot lines such as \fe.

\section{Ray Selection }\label{raysel}

\subsection{ White Light Sample}

The white light ray-like features we analyzed in this paper were selected from the LASCO C2 observations. 
A catalog of LASCO CMEs with trailing concave-outward structures and rays has been developed by Webb. 
A comprehensive paper on the results of analysis of the WL rays is in preparation (Webb 2012, in preparation). 

\begin{figure*}[h]
\centering
\begin{tabular}{cc}
  \includegraphics[width=6cm]{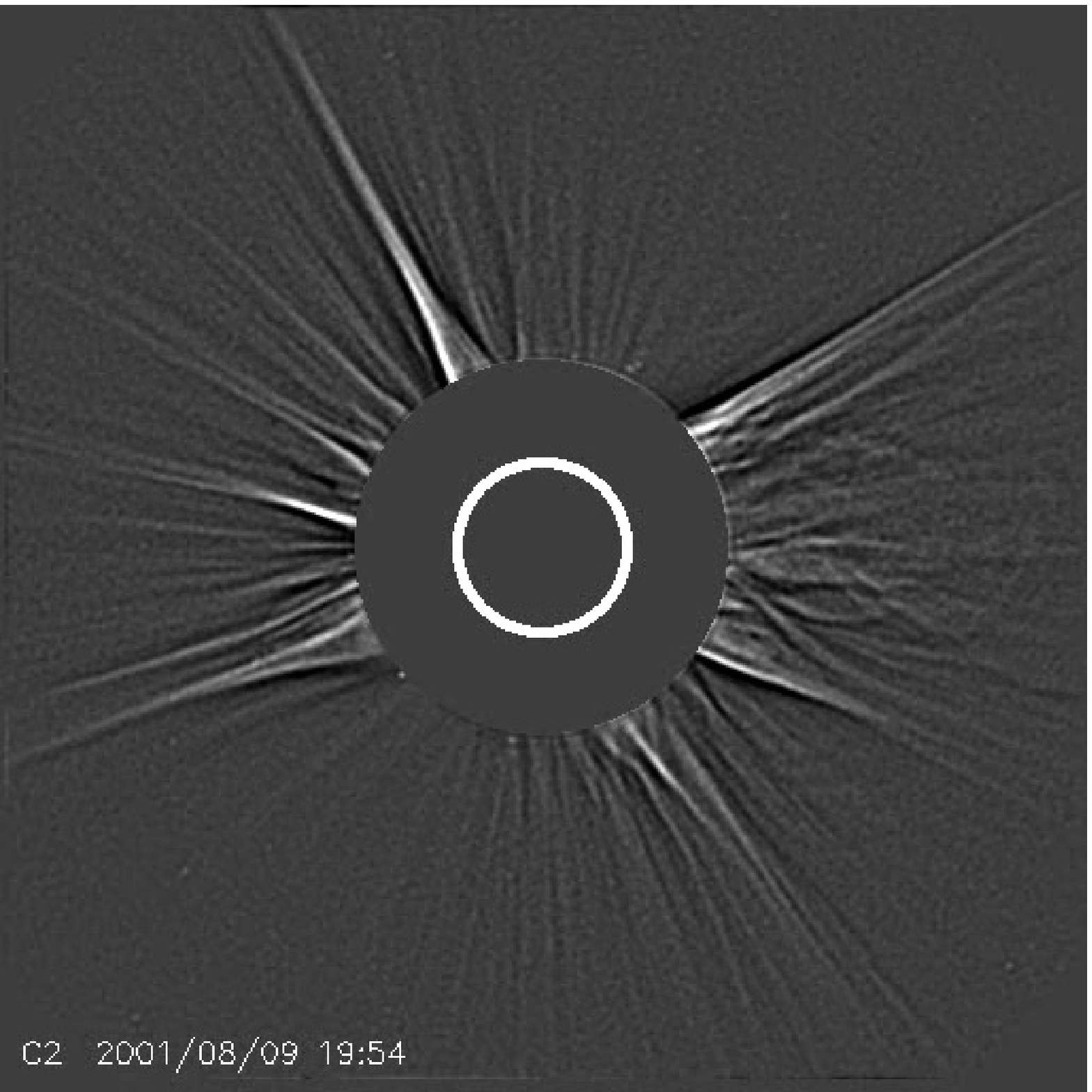} &
  \includegraphics[width=6cm]{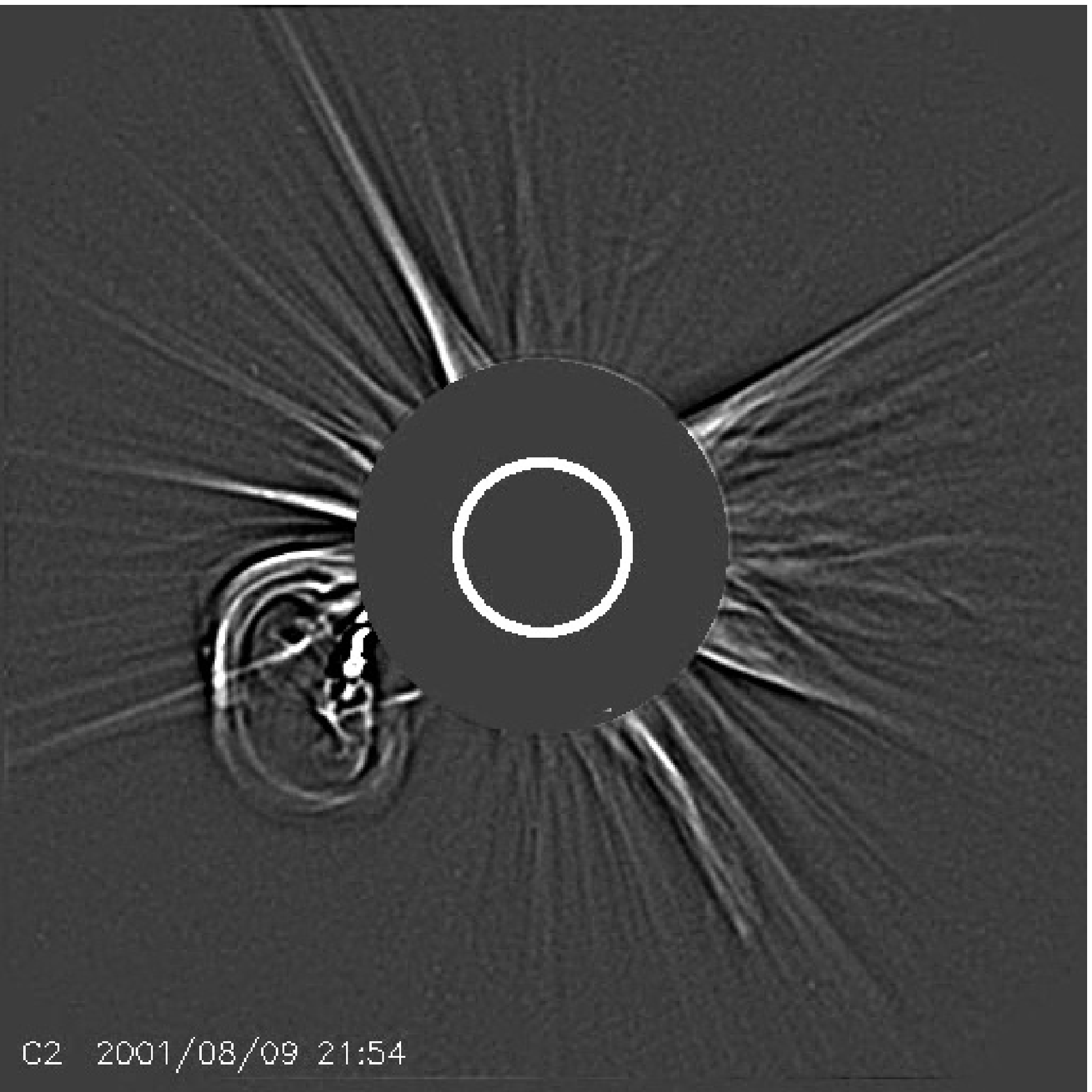} \\
  \includegraphics[width=6cm]{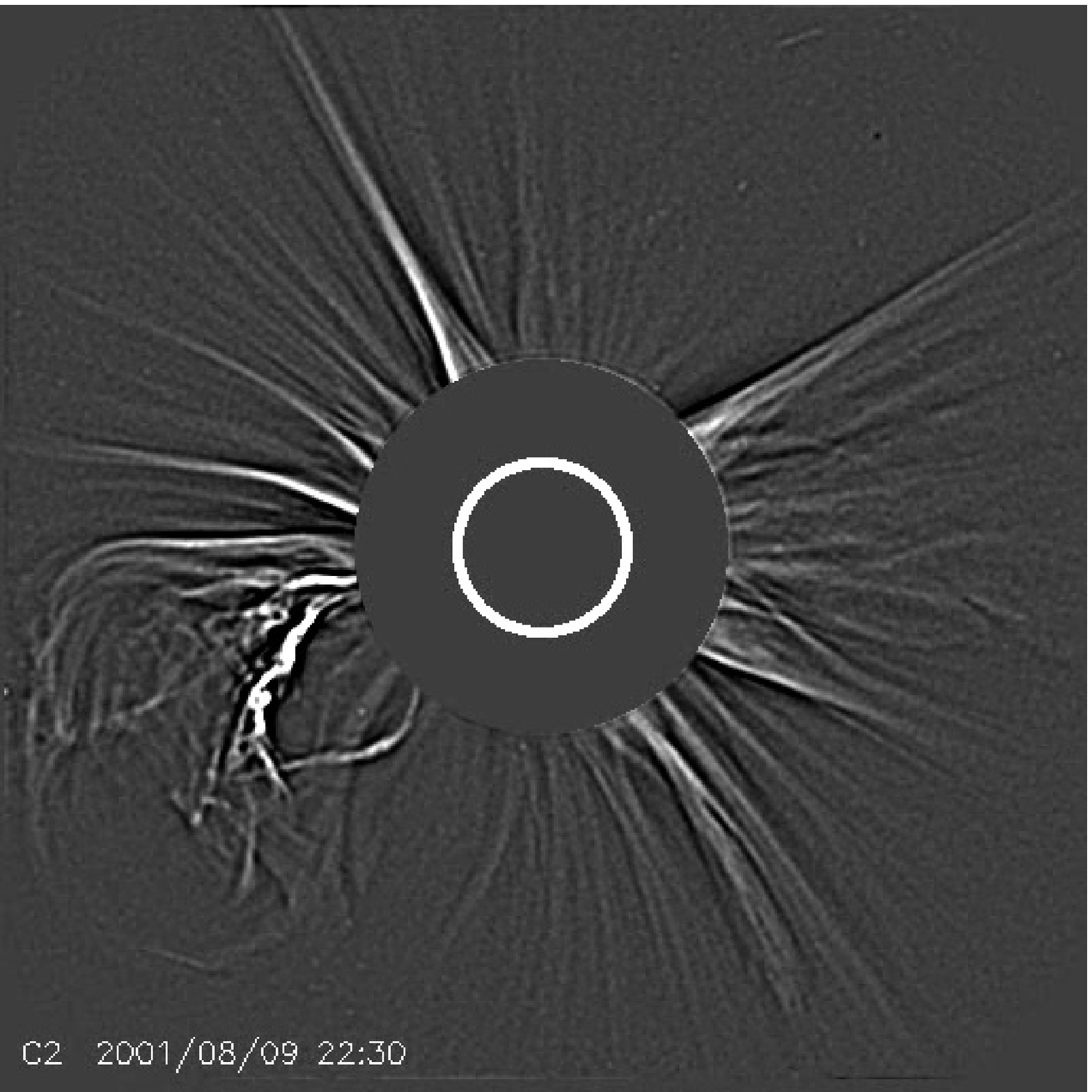} &
  \includegraphics[width=6cm]{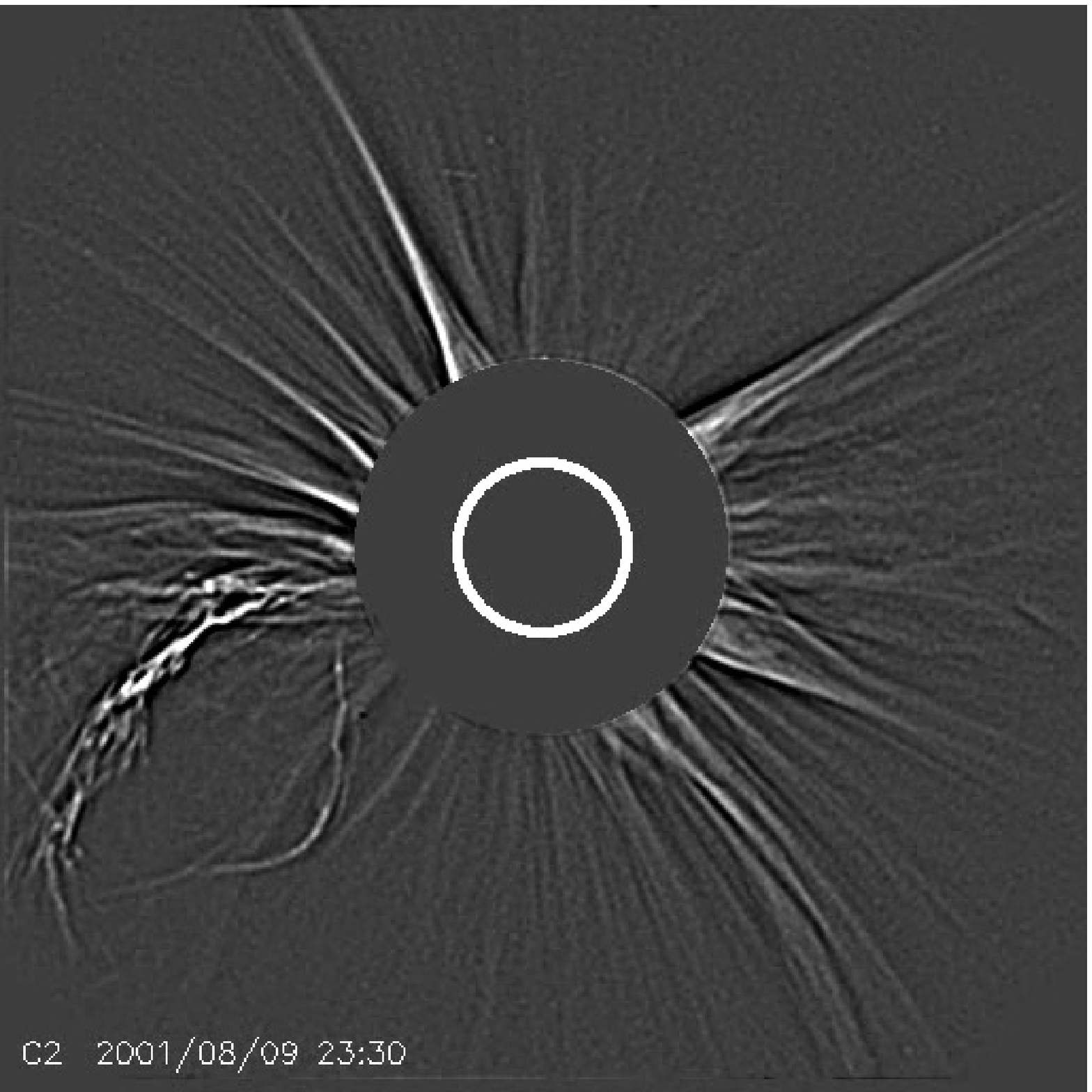} \\
  \includegraphics[width=6cm]{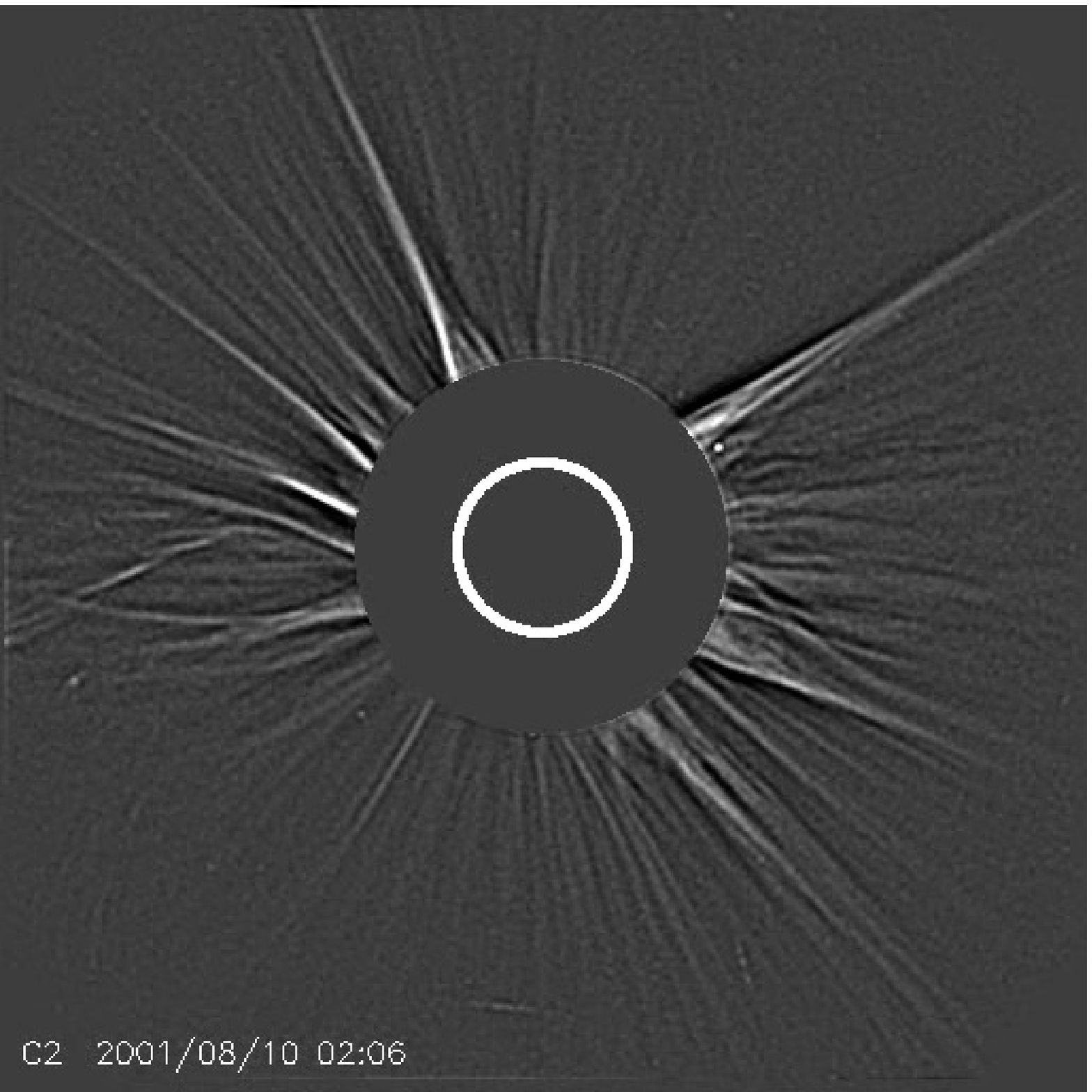}&
  \includegraphics[width=6cm]{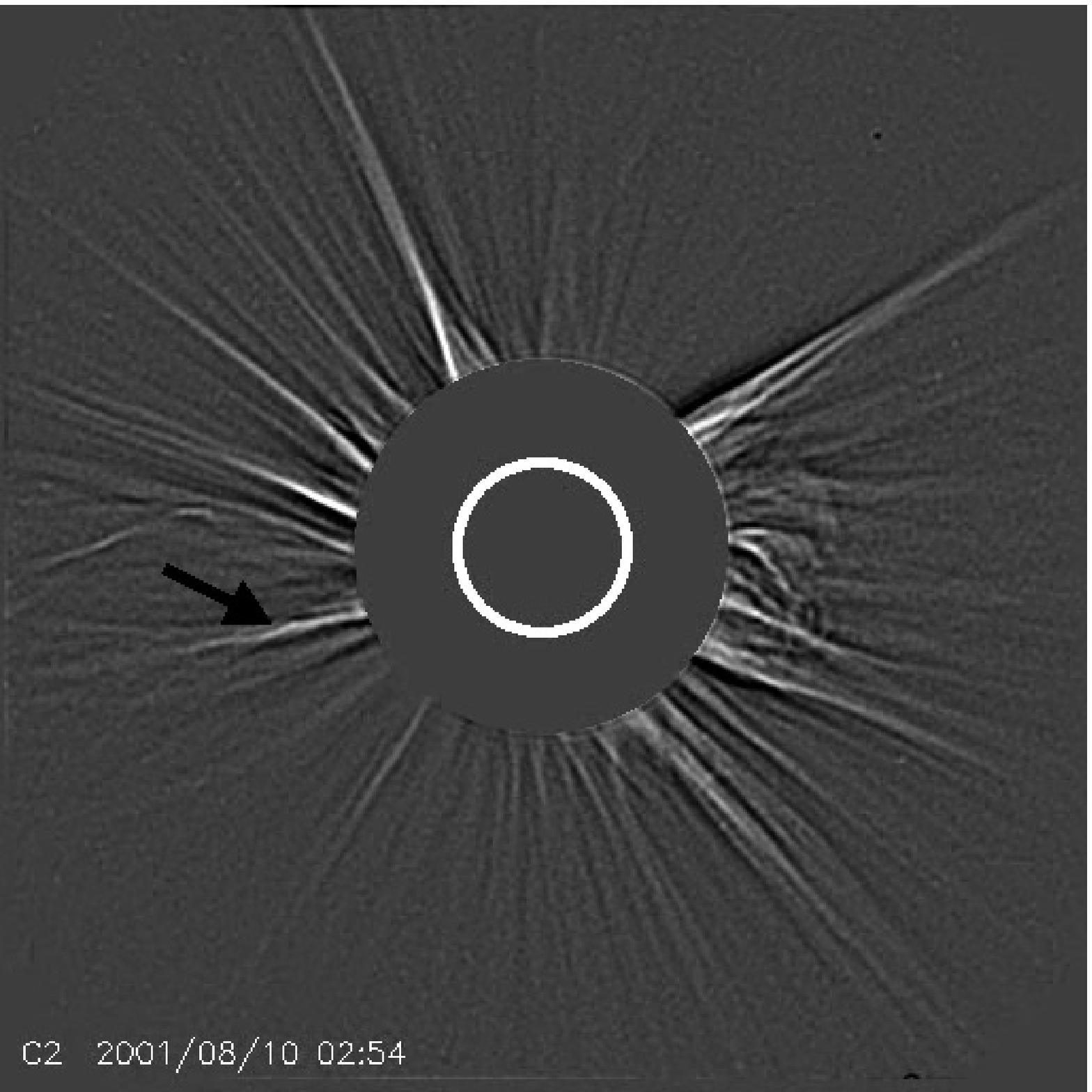} \\
  \end{tabular}
  \vspace{-0.2cm}
   \caption{LASCO images of the CME on 2001 August 9. The images show the 
   pre-CME corona (top left) at 19:54 UT and the CME evolution at 21:54, 22:30 and
   23:30 UT. The bottom row shows the appearance of the ray at 02:06 UT and the clearly formed 
   ray at 02:54 UT.}
   \label{lasco_ray}
\end{figure*}

Early studies found that about 20\% of the CMEs showed
concave-outward features in SMM and LASCO coronagraph data, and these were identified as magnetic disconnection 
events (\citealt{W95,W03}). \cite{W95} examined SMM coronagraph data for bright rays in the wakes of CMEs that were 
coaxial with the CME and had outward-moving U- or V-shaped structures at the top. Recently, \cite{hdr12} have presented
beautiful examples of a few disconnection events seen in STEREO coronagraph data and tracked their evolution into the heliosphere. 
In Figure~\ref{lasco_ray} the sequence of LASCO wavelet-processed images shows the CME on 2001 August 9 moving outwards and the bright ray behind it. 
During this event the pre-existing streamer has blown out and on August 10 at 02:06 UT 
the transient ray has started to form. This is an event wherein the ray becomes a new, 
reformed, streamer. The depletion of \Si12 emission, as well as \LA and \Osix  in the region of the pre-CME streamer supports the LASCO movie interpretation of a ray-reforming streamer. 

\begin{figure*}[ht]
\centering
   \includegraphics[width=16.5cm]{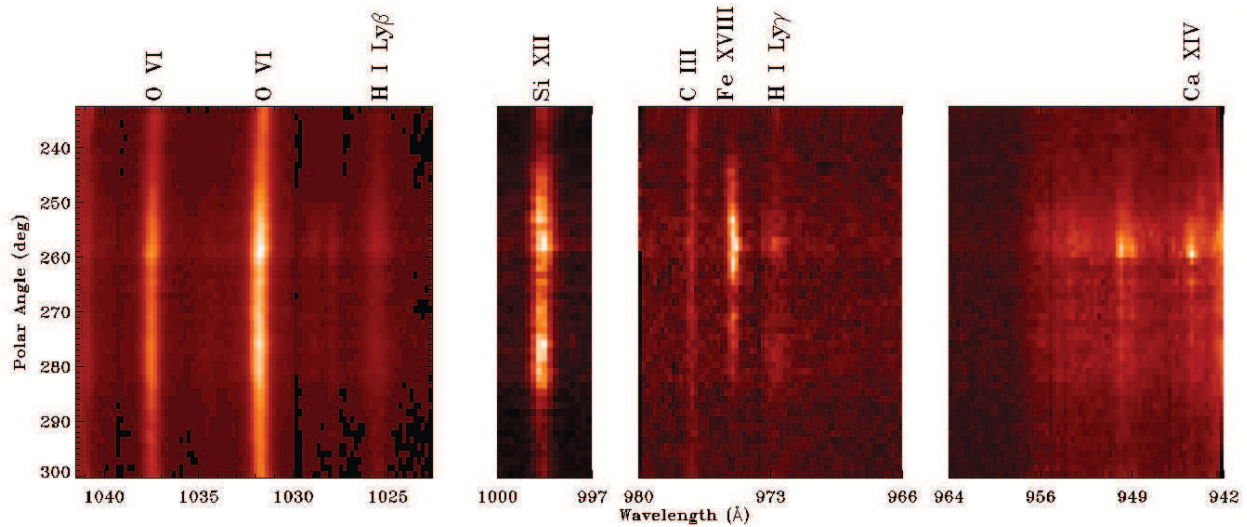}
   \caption{UVCS spectrum observed on 2003 June 2 after a CME. Some of the main lines are marked. The 
second order lines of \Si12 \L521 is the first line on the left of first panel, and \L499 is in the second panel.}
   \label{uvcs_cs}
\end{figure*}

The selection criteria for the events in the WL catalog are essentially those described by \cite{W03} and summarized above. Events were selected that had a bright, 
narrow, linear ray following the concave-outward structure and coaxial within a few degrees with the CME 
volume. We required that such a ray appear within no more than about 12 hours following CME onset. Adequate data coverage was required during this period, ideally for at least a full day starting from just 
before CME onset to enable the identification of pre-existing structures, especially streamers, and whether 
a new streamer appeared in the wake of the CME and any ray. 
Indeed many of the WL rays form after streamer blowout and are associated with the subsequent reformation of 
the streamer \citep{W03}. In our sample 29 events ($\sim$ 40\%) are associated with streamer 
blowout, and in 7 cases the streamer forms afterward.
For the events selected, we recorded the time when the ray appeared, its position 
angle and duration, along with the parameters of the CME. If Mauna Loa data for the appropriate
time period were available, they were examined for the presence or absence of a corresponding ray. 
The ray-like features analyzed in this paper were selected from the LASCO ray catalog, mainly within the 
two temporal windows: from July 1996 to Dec 1998 during solar minimum, and 
2001 for solar maximum.  A total of 157 rays have been selected from LASCO images, of which 
74 occur during the solar minimum period and 83 during the maximum.
Once the white light rays were selected, UVCS archival data was examinaexaminedted for the 
existence of spectra at the appropriate position angle and time.
There were cases in which UVCS did not observe any 
narrow bright feature corresponding to the WL ray. 
Of the 157 LASCO rays, 60 ($\sim$ 47\%) had appropriate
UVCS spectra showing bright features that matched the WL rays. 
Of these, 26 were found during the solar minimum period
and 34 during solar maximum.   
Narrow bright features in the H I, O VI, Si XII and if possible
\fe lines were searched for at the position angle of the white light feature. 
Some misalignment in position angle was tolerated because the features evolve in 
time and are not always radial,
and most of the UVCS observations covered regions below the LASCO occulter height of 2.3 
\RSUN.  In general,
if a UVCS feature was found, it was within about 3\DEG\/ of the LASCO position angle,
but we include a few events with larger position angle differences.
Most of the rays were detected during the daily synoptic scans,
which unfortunately used a narrow spectral range and often did not include \fe line.
We chose 2001 for this study because the synoptic sequence did include the \fe line.  

\subsection{UVCS Sample}

In the wake of CMEs UVCS has observed several spatially narrow and bright features in 
the hot line of \fe that have been interpreted as signatures of the current sheet (CS) 
associated with CMEs (\citealt{Cia02,Ko03,Be06,Cia08}).  Figure~\ref{uvcs_cs} shows the spectrum of the 
CS associated with the CME on 2003 June 02 at 08:54 UT. The narrow bright feature in \fe appeared 
after the CME passed the UVCS field of view, and it was well aligned with the prominence core. 
This feature was also detected in the [Ca XIV] line in the fourth panel.

UVCS has observed more than a thousand CMEs and a catalog \citep{Gio13}, 
linked to the CDAW catalog of LASCO CMEs ($\rm http://cdaw.gsfc.nasa.gov/CME\_list/$)
was constructed by examining the UVCS data for each LASCO CME from the start of 
regular SOHO operations in 1996 through the end of
2005.  For each white light CME, a special purpose computer code selected UVCS data at the 
position angle of the CME within 4 hours of the event. 
As part of the catalog, we determined whether there was an indication of a current sheet,
usually by the presence of a narrow feature in \fe. 
For the present analysis, we examined the
spectra of the 18 current sheets that showed clear features in \fe and compared them with LASCO
data to determine whether a WL ray with current sheet morphology coincided
with the UV feature.
The sample includes the 6 events for which analysis of the current sheet has been 
previously published.

\begin{figure*}[ht]
\centering
\begin{tabular}{cc}
   \includegraphics[width=6.0cm]{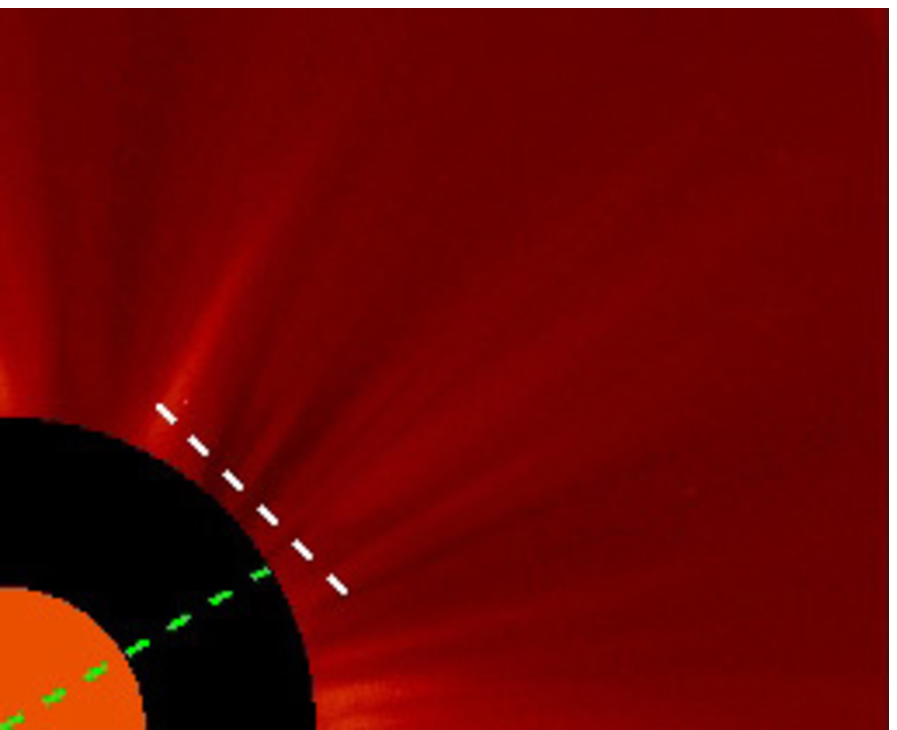} &	\includegraphics[width=6.0cm]{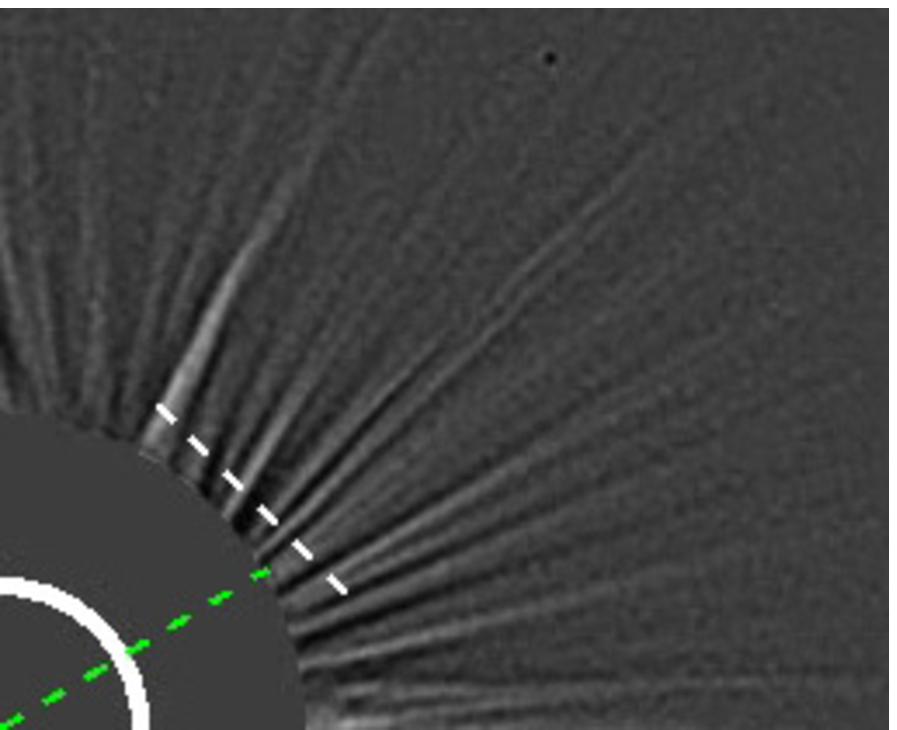}\\
   \includegraphics[width=6.0cm]{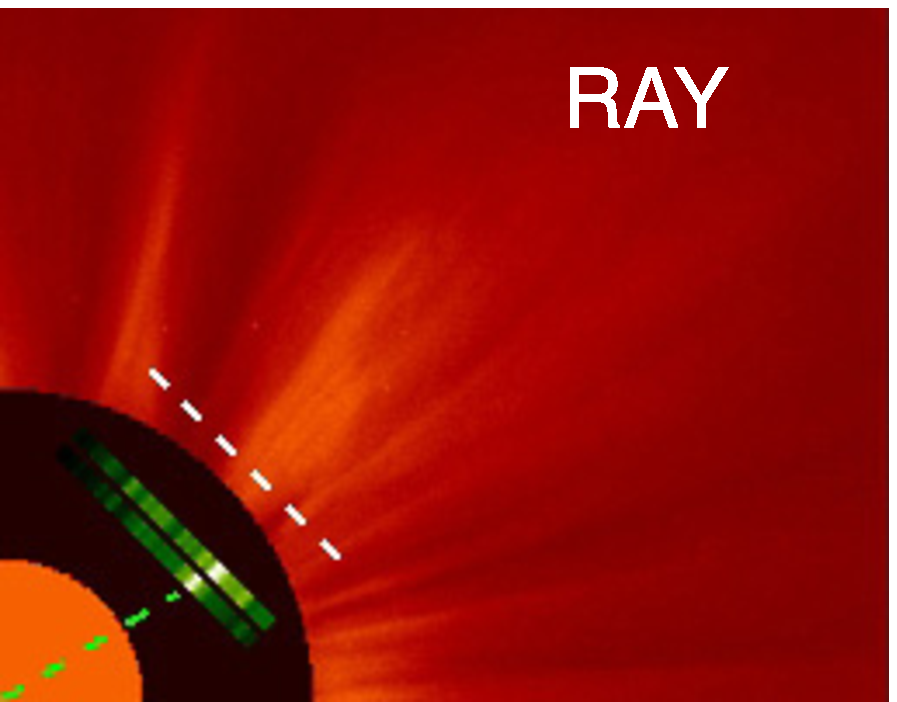} &	\includegraphics[width=6.0cm]{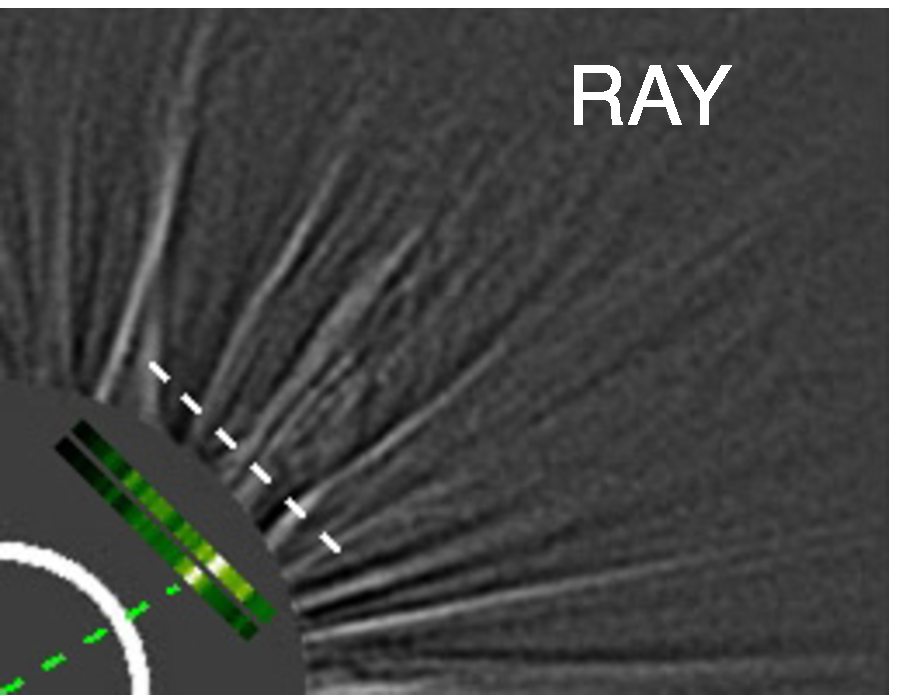}\\
   \includegraphics[width=6.0cm]{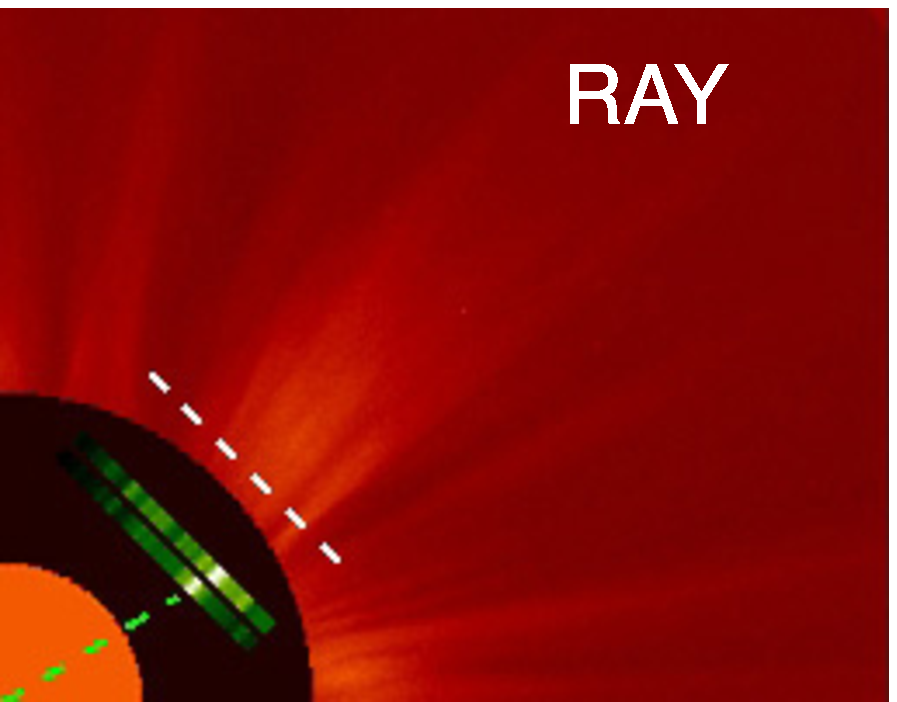} &	\includegraphics[width=6.0cm]{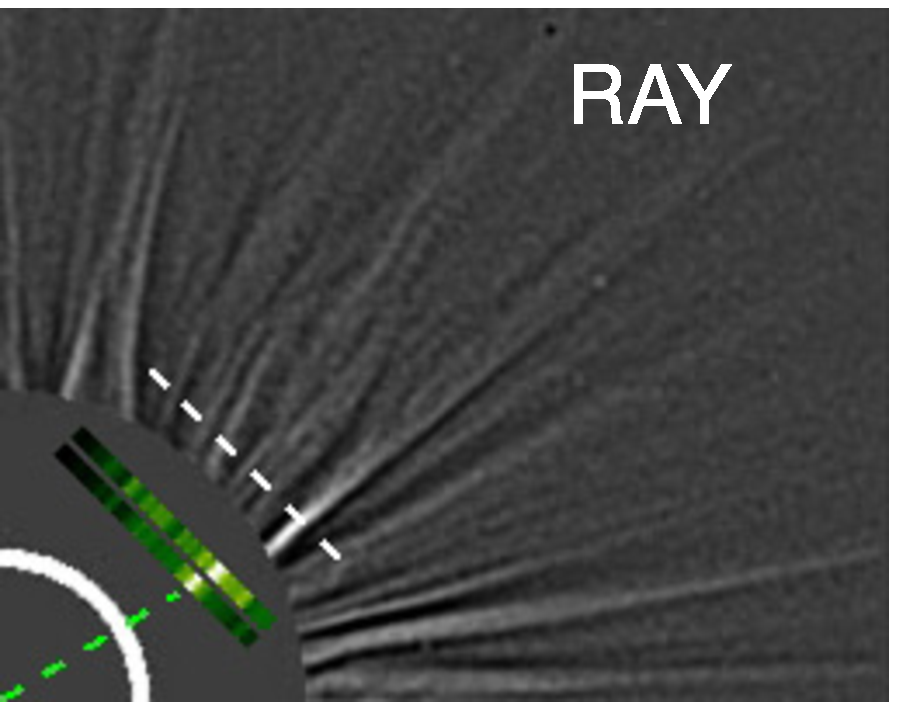}\\
   \includegraphics[width=6.0cm]{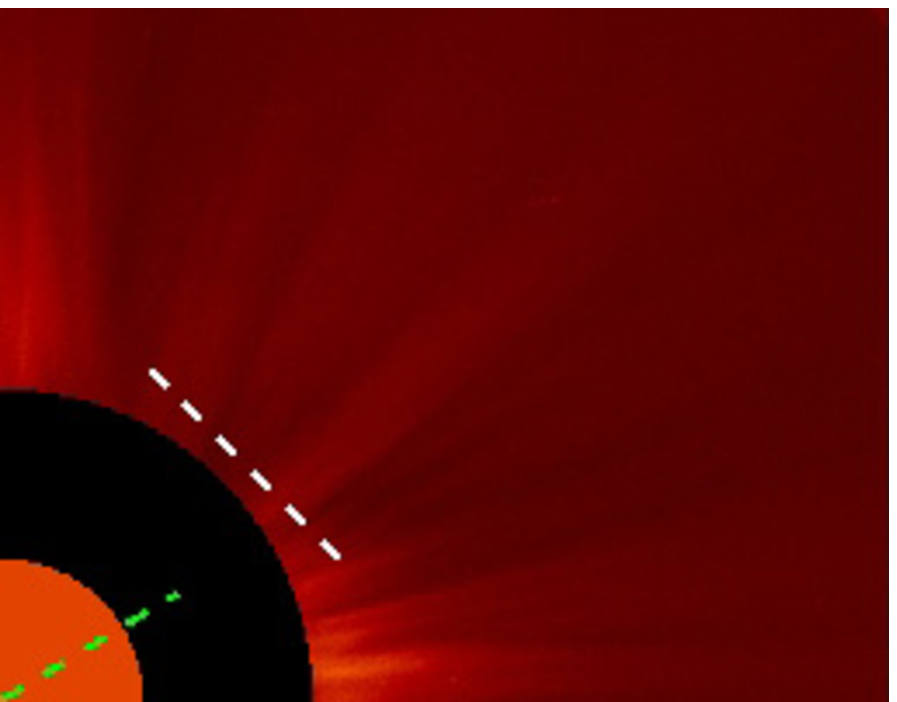} &	\includegraphics[width=6.0cm]{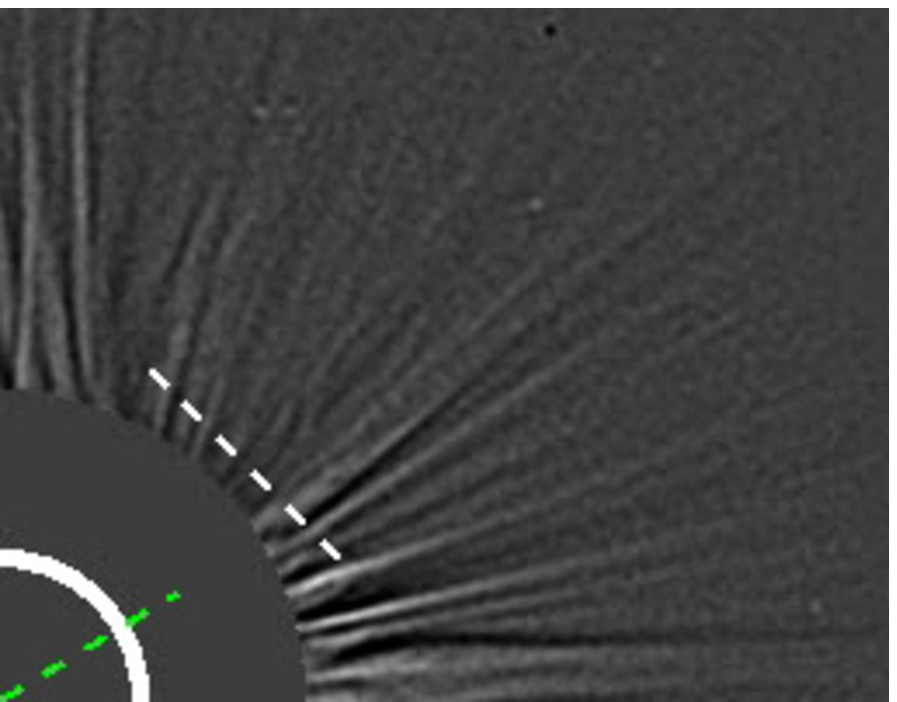}\\
  \end{tabular}
 \caption{LASCO C2 images and wavelet-processed of the ray associated to the CME on 2000 September 23 at 21:30. The images show the 
pre-CME corona (top row) on 23rd at 20:26 UT, the ray in the wake of the CME (2nd and 3rd rows) on 24th at 01:27 and 10:26 UT 
and the post-ray corona (bottom row) on 24th on 23:50 UT. The green dashed line 
indicates the position of the WL ray.  The images below the occulter are \fe along the UVCS slit taken on 23rd at 1.67 \RSUN (from 22:19 to 22:28 UT) and at 1.53 \RSUN (from 22:29 to 22:33 UT). The white dashed line at 2.4 \RSUN 
marks the area where the LASCO data have been extracted to obtain the intensity distribution for Figure~\ref{pl_000923}.}
   \label{c2_000923}
\end{figure*}

\section{Ray Analysis}\label{anal}

\begin{figure*}[ht]
\centering
\begin{tabular}{cc}
   \includegraphics[width=8cm]{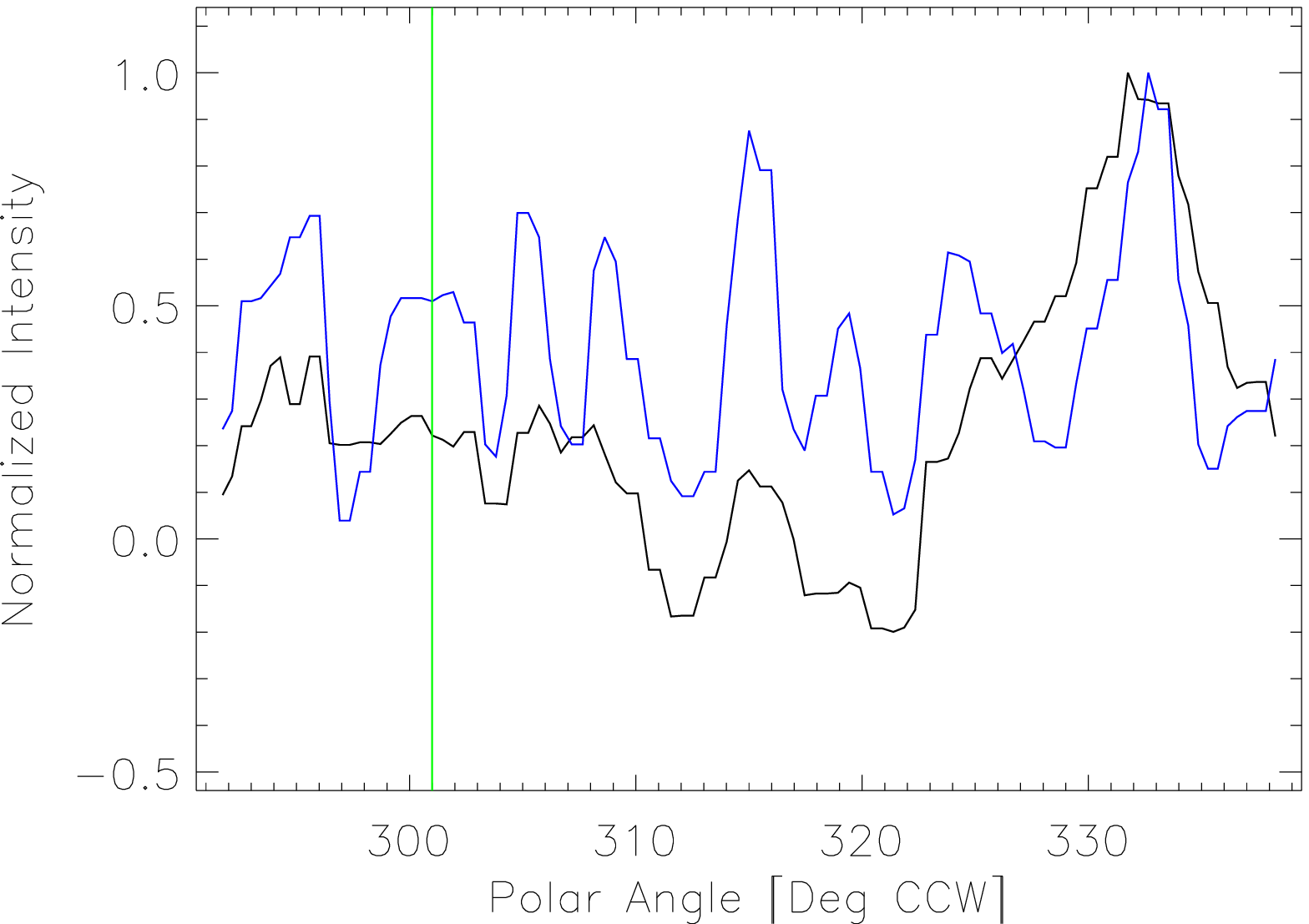} & 
   \includegraphics[width=8cm]{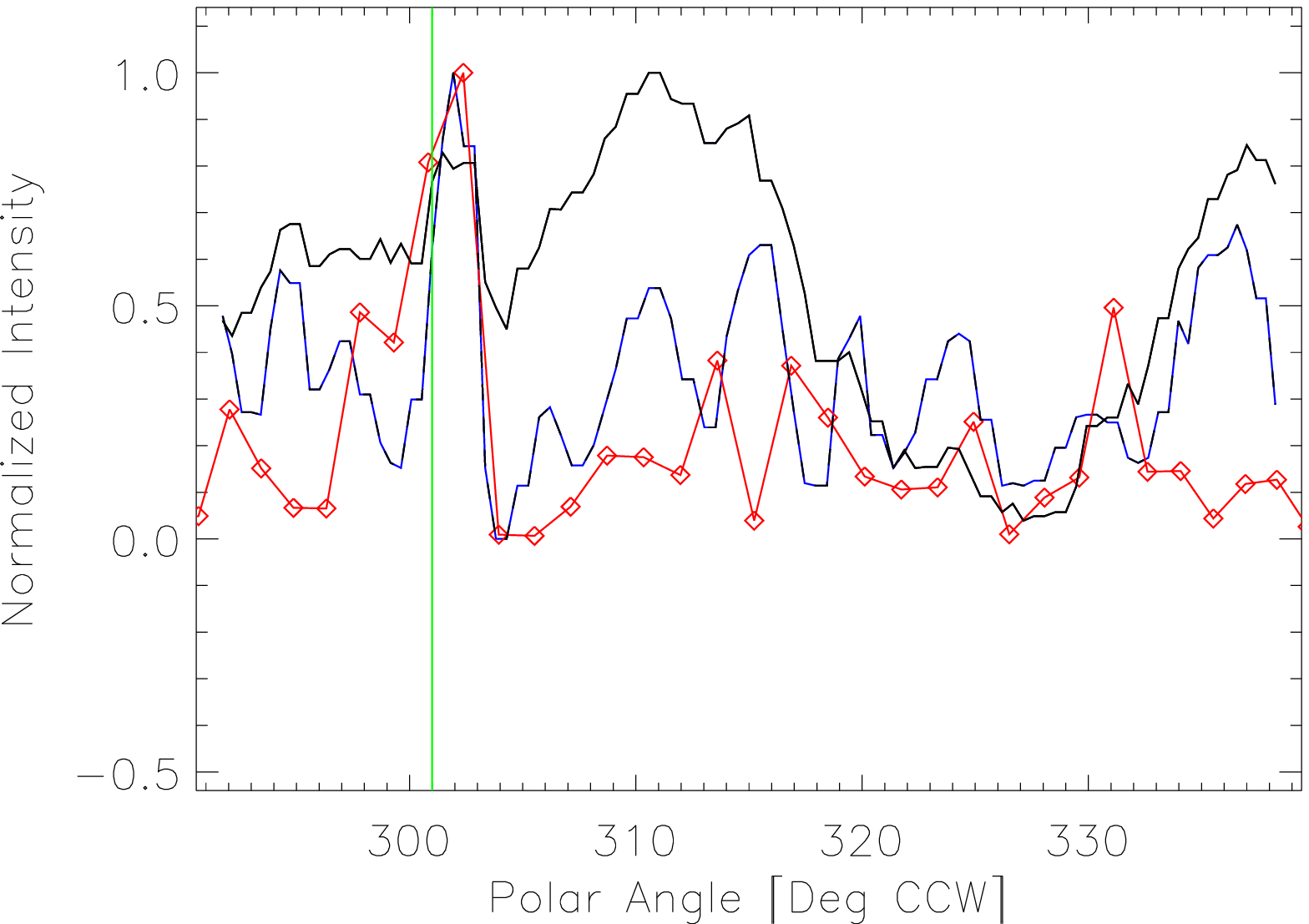} \\
   \includegraphics[width=8cm]{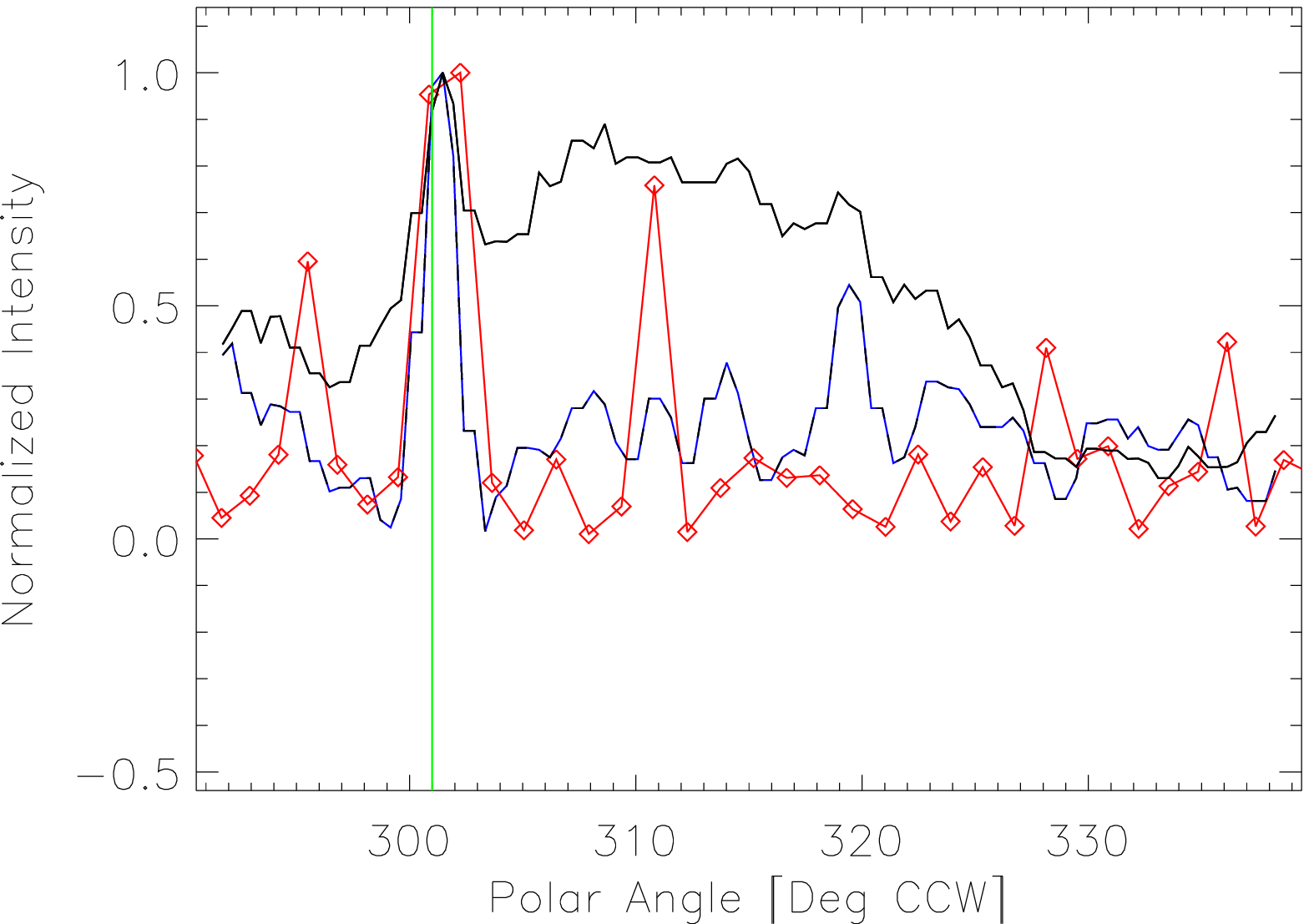} &
   \includegraphics[width=8cm]{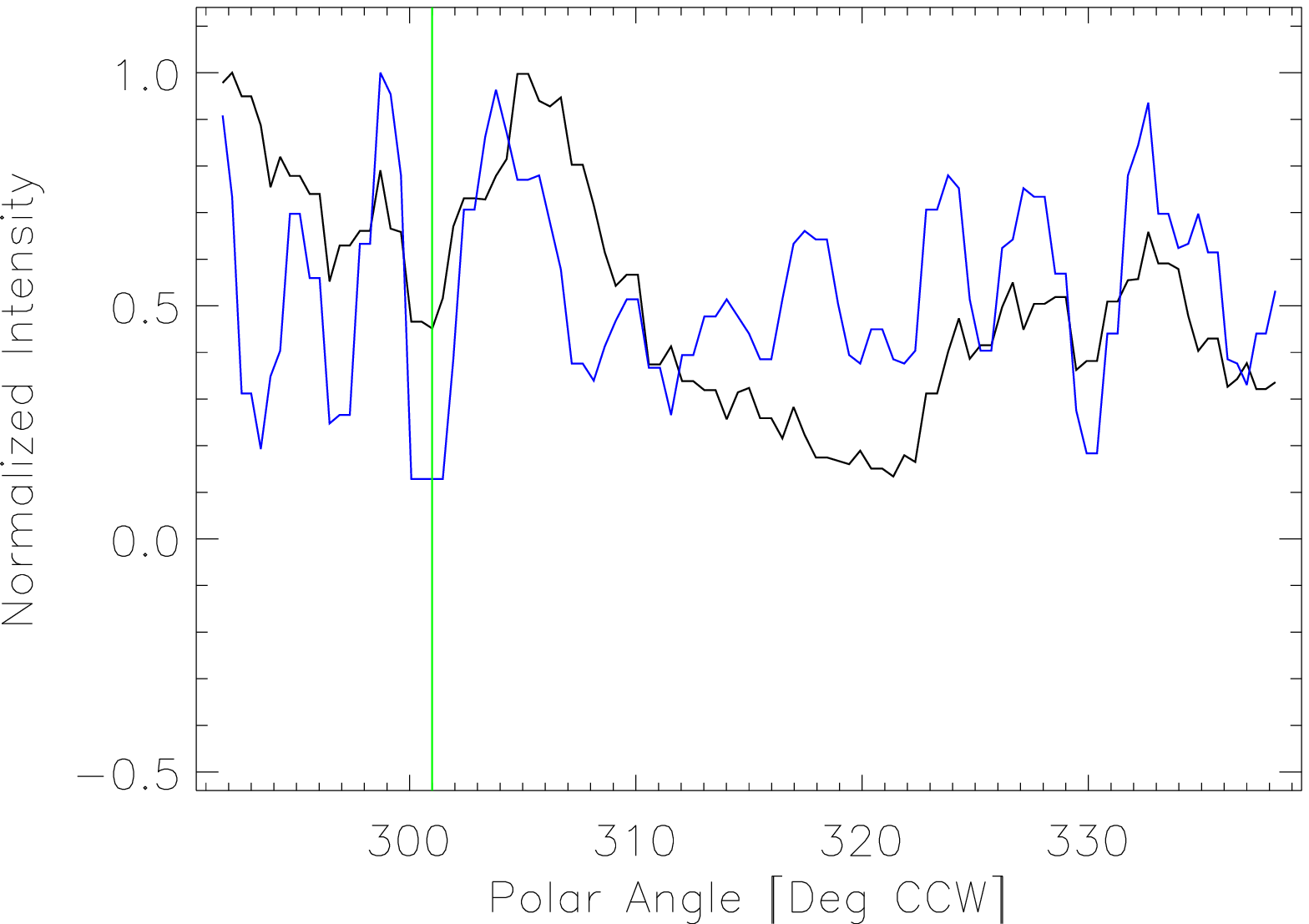} \\
  \end{tabular}
  \vspace{-0.2cm}
   \caption{The four panels show the normalized intensities of LASCO C2 at 2.4 \RSUN along the white segment in Figure~\ref{c2_000923}. The green line marks the position of the ray in the WL images.
Black and blue line correspond to LASCO background subtracted and wavelet-processed images respectively.
The red line is the \fe line intensity along the UVCS slit at 1.67 (top right panel) and 1.53 \RSUN
(bottom left panel).}
   \label{pl_000923}
\end{figure*}

In our analysis we used LASCO WL background subtracted and wavelet-processed images.
The background subtracted images have been obtained subtracting from LASCO raw images
the related monthly minimum background and removing the spikes mainly due to the cosmic
rays.
The wavelet technique is particularly suitable for analysis of data containing 
discontinuities or sharp spikes and makes it possible to recover weak signals from noise. 
It uses mathematical functions to decompose data into different frequency components 
and then study each component with a resolution adapted to its scale. 
The wavelet-processed images from LASCO archive are based on a 
techniques developed by \cite{ste03}.

Before presenting our results we will describe the analysis method used for all the rays reported 
in this work by discussing in detail three events representative of  hot, coronal and cool rays. 
We define hot as the rays that show \fe emission, coronal as those that are detected in 
\Si12 and \Osix, and cool rays 
as those that show emission only in the lines  of \Osix and/or \LA. 
We note that \Osix emission peaks at a temperature of about 0.3 MK, but that it persists
up to about 2 MK due to dielectronic recombination of O VII.  \LA also peaks at low
temperatures, but the combination of large hydrogen abundance and strong illumination
by chromospheric \LA photons combine to make it the strongest coronal UV line in spite
of the $10^{-6}$ H I fraction.  These lines seen in combination with coronal lines
such as \Si12 indicate a temperature similar to that of the corona.  Bright emission
in this lines without corresponding coronal emission lines indicates a temperature below
1 MK.  This is often corroborated by very cool lines such as \C3 and by very narrow 
line widths.  

{\it Hot Ray (2000 September 23)} -
The WL images of the event have been analyzed before, during and after the CME
in order to detect rays with the characteristics described in the previous paragraph.
Figure~\ref{c2_000923} shows a sequence of LASCO and related wavelet-processed  
images for the event on 2000 September 23.
>From the movies, images of the pre-CME corona (top row), of the ray in the wake of the CME
(2nd and 3rd rows), and of the post-ray corona (bottom row) have been selected. The dashed 
green line in the images marks the position of the ray, the images below the occulter are 
the \fe images along UVCS slit obtained at 1.53 and 1.67 \RSUN. 
Both background subtracted and wavelet-processed images show clearly the formation 
of a new ray. The bright feature along the UVCS
slit is well aligned with the WL ray. The pre-CME and post-ray coronal images in the \fe line do not show 
any sharp emission at the location of the ray. In Figure~\ref{pl_000923} we compare the normalized LASCO intensities along the white 
segment marked in Figure~\ref{c2_000923} with the normalized intensity of \fe (red line) along the UVCS slit at 1.53 \RSUN (top right panel) 
and 1.67 \RSUN (bottom left panel).
The ray was not detected in the in \LA, \Osix or \Si12 lines, see Table~\ref{fe_ray}. \fe is the only line in which we detected a bright peak at the 
the location of the WL ray, and this implies a temperature of 10$^{6.8}$ K or higher.
The black and blue lines correspond to the background subtracted and wavelet-processed LASCO 
images, respectively.
While the LASCO intensities are extracted at 2.5 \RSUN and the position of the  
UVCS entrance slit is below the LASCO occulter, the WL and UV ray are in good agreement. The UV ray at 1.53 \RSUN is
offset toward north by about 2\DEG, probably because the ray is not exactly radial.

\begin{figure}
\begin{tabular}{cc}
\centering
   \includegraphics[width=3.75cm]{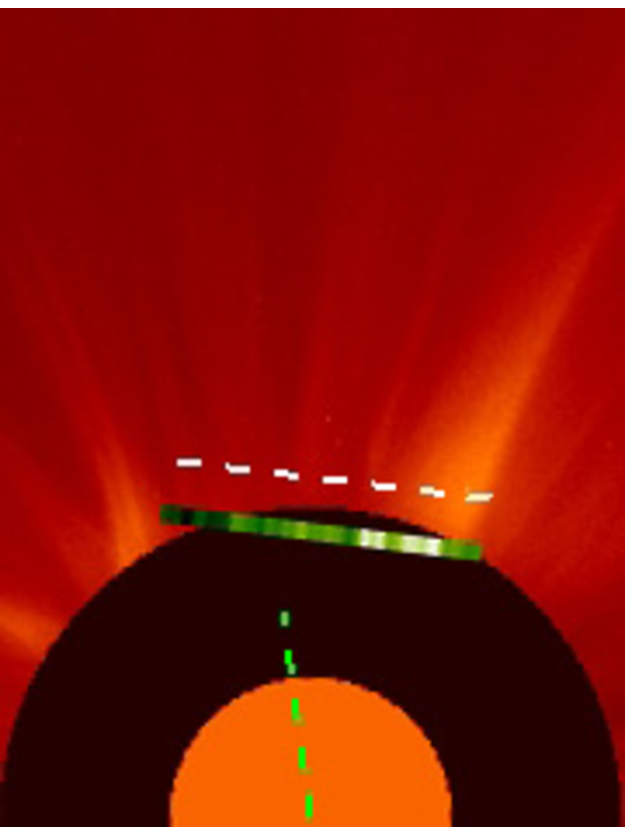} &	\includegraphics[width=3.75cm]{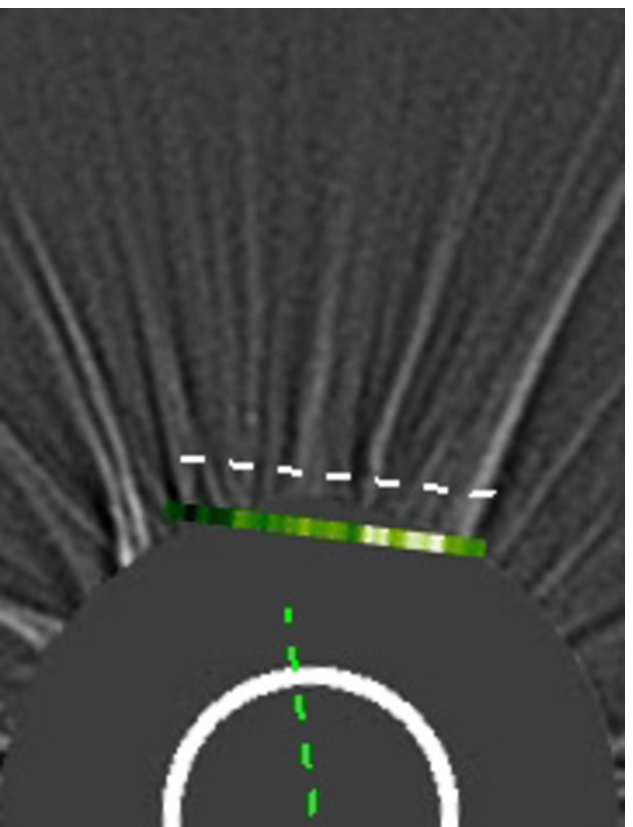}\\
   \includegraphics[width=3.75cm]{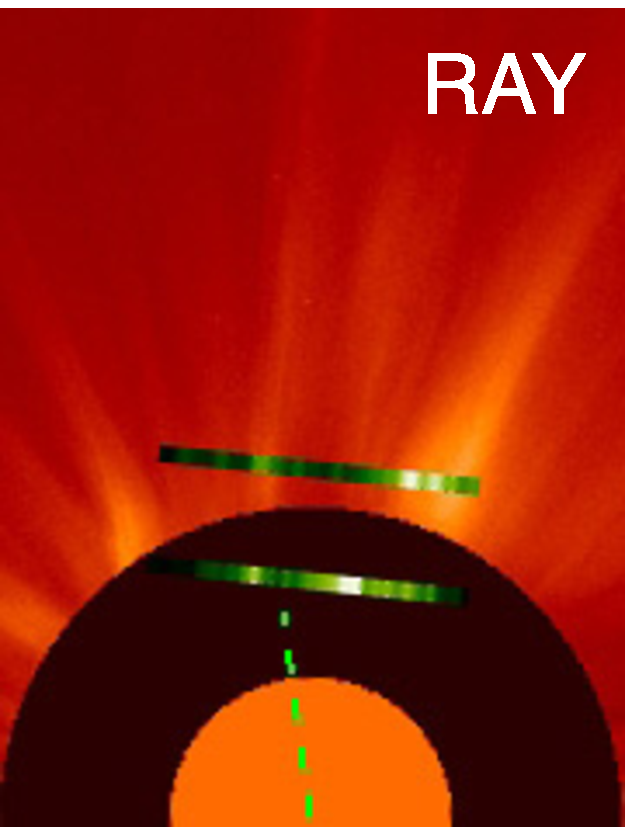} &	\includegraphics[width=3.75cm]{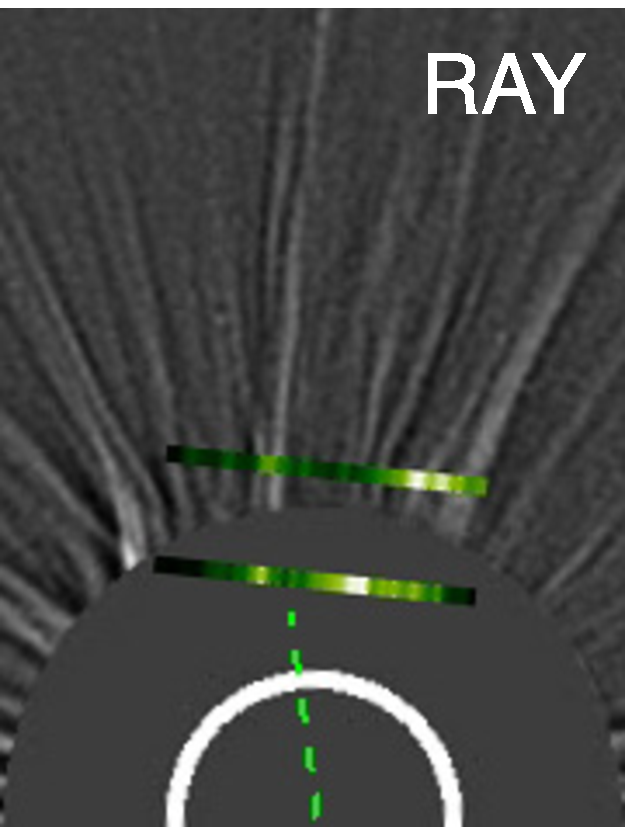}\\
   \includegraphics[width=3.75cm]{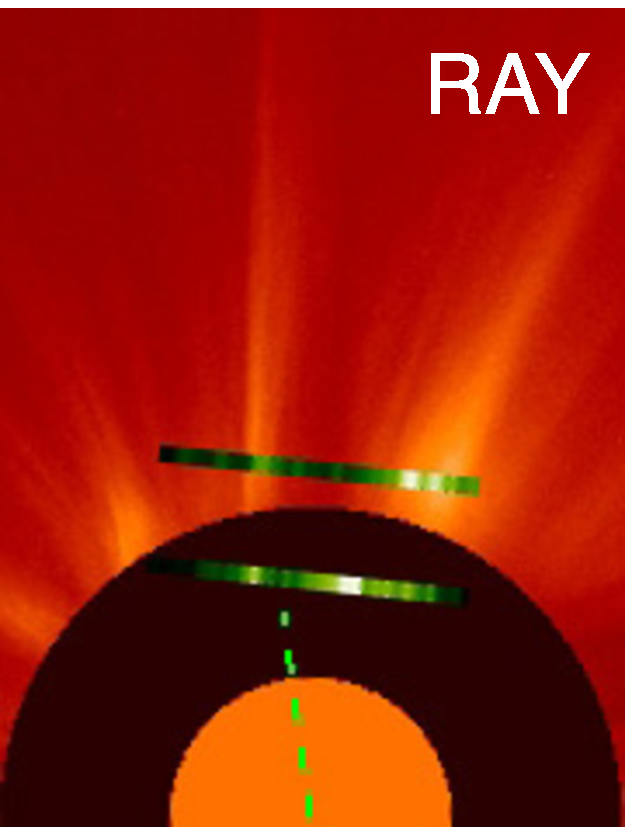} &	\includegraphics[width=3.75cm]{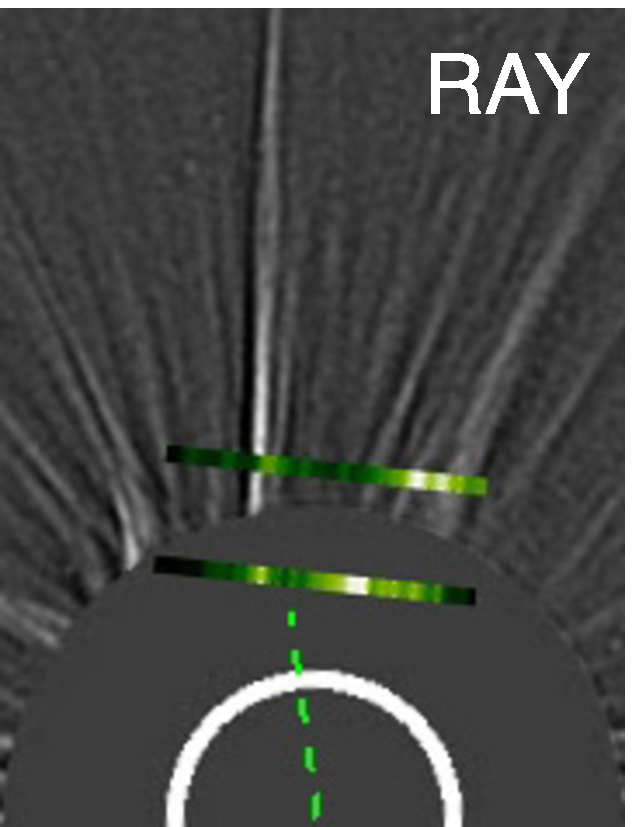}\\
   \includegraphics[width=3.75cm]{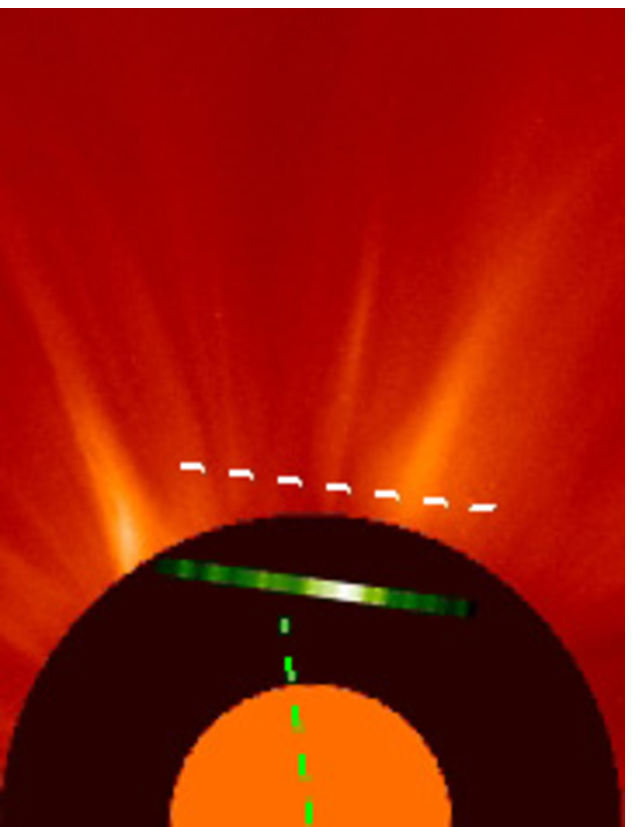} &	\includegraphics[width=3.75cm]{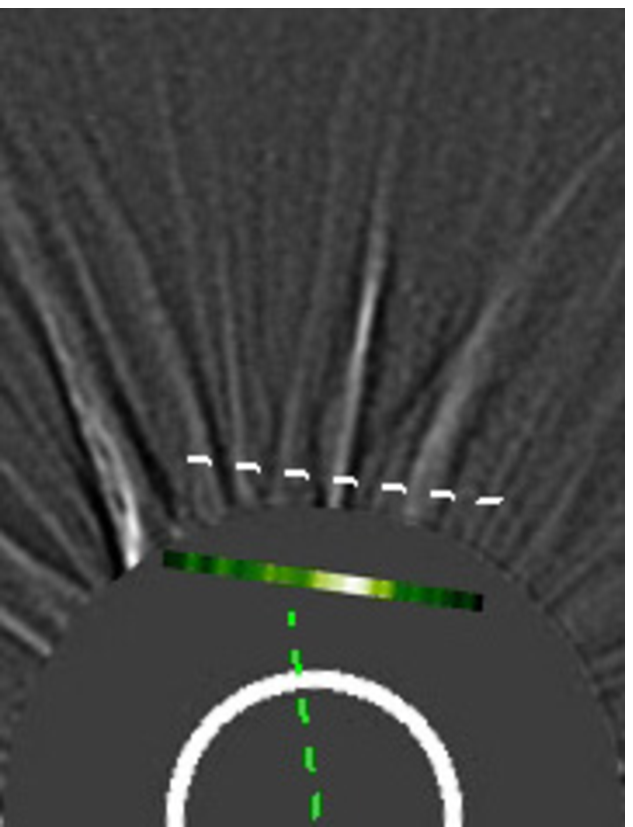}\\
\end{tabular}

   \caption{LASCO C2 images and wavelet-processed of the ray associated to the CME on 2001 May 12 at 23:00 UT with  
     superimposed the Si XII 499\AA\ intensity image along the UVCS slit. From the top the sequence shows the pre-CME corona 
     on 2001 May 12 at 22:15 with Si XII image from 21:34 to 22:25 UT at 1.97\RSUN, 
     two images of the ray on 2001 May 13 at 02:41 and 05:50 UT in LASCO C2 and Si XII at 2.42\RSUN, from 03:48 to 05:31 UT. The bottom images 
     show the post-ray corona in LASCO C2 at 19:30 UT and UVCS Si XII, at 1.67\RSUN, from 19:28 to 20:17 UT. The white dashed line 
     marks the position at which the LASCO data have been extracted when the UVCS
     slit was located below the occulter disk.}
   \label{c2_010512}
\end{figure}

\begin{figure*}
\centering
\begin{tabular}{cc}
  \includegraphics[width=8cm]{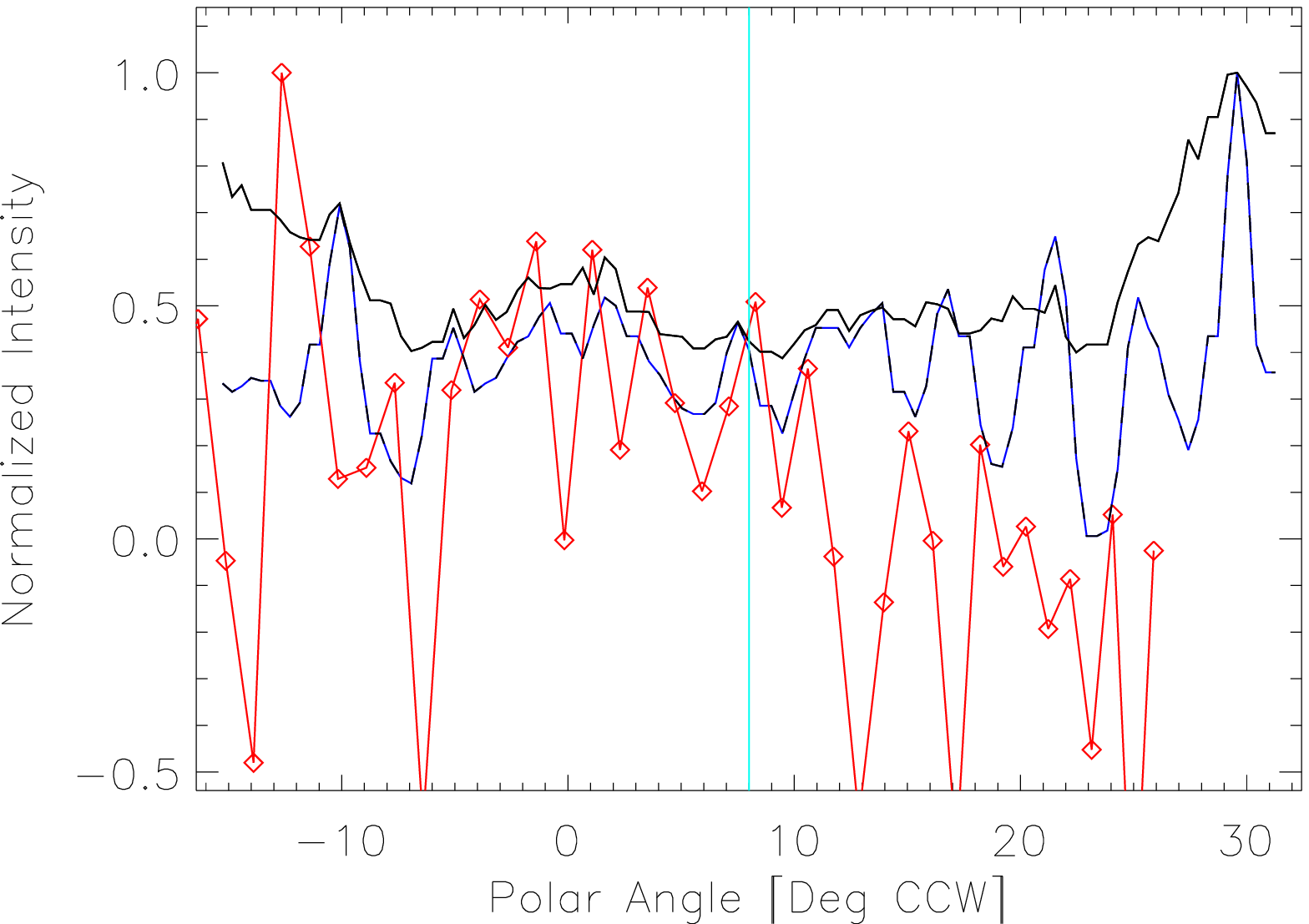} &
  \includegraphics[width=8cm]{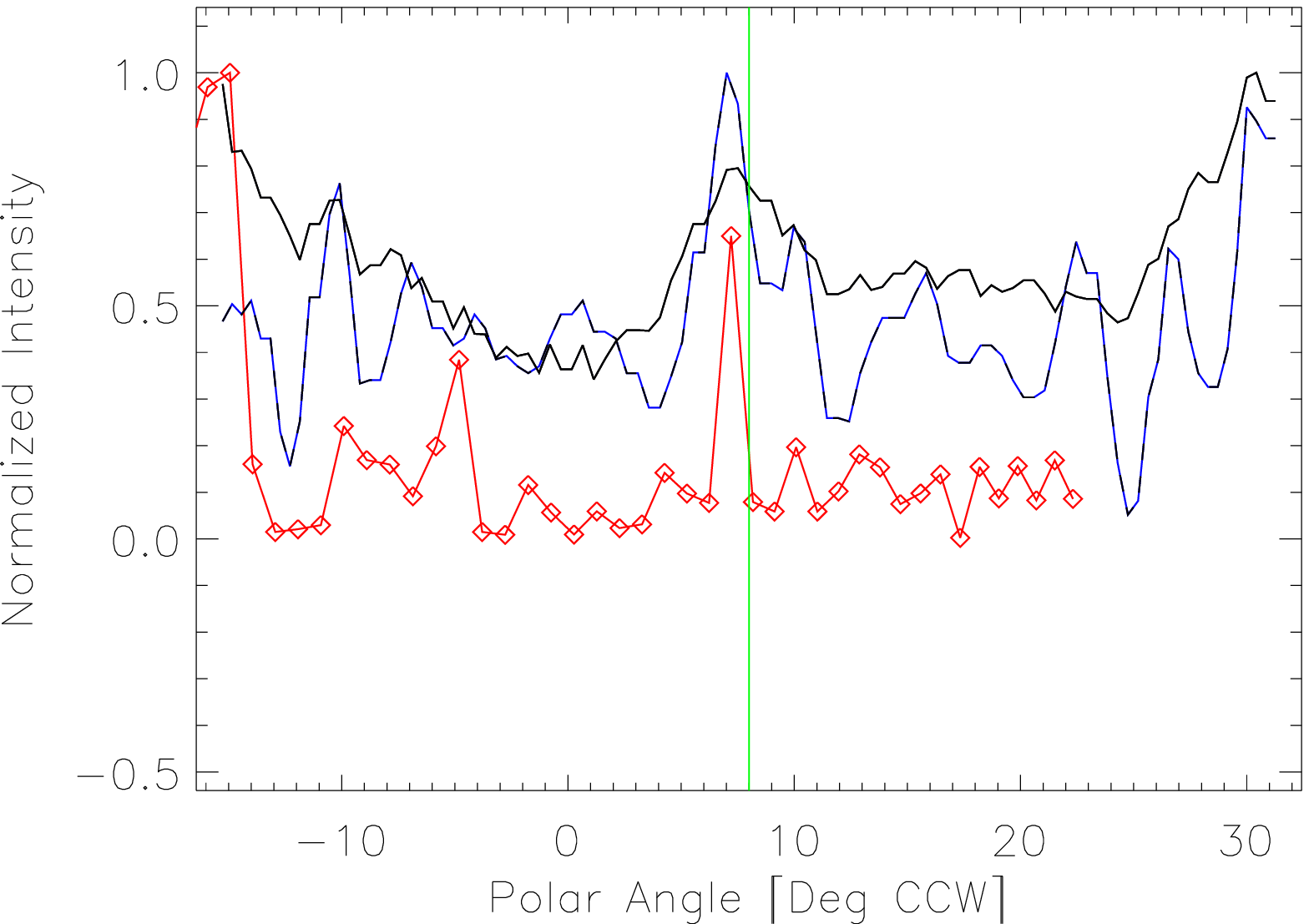} \\
  \includegraphics[width=8cm]{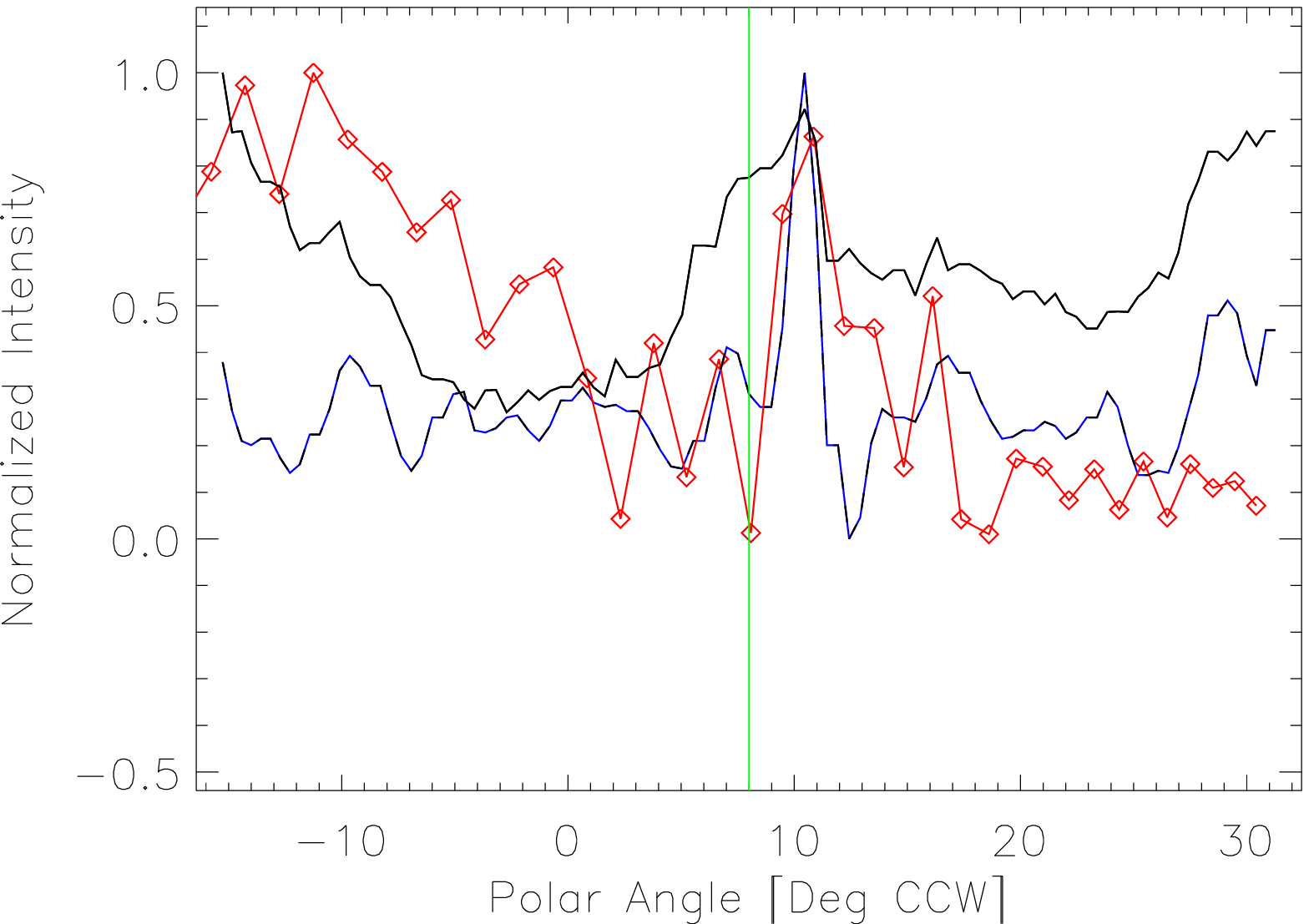} &
  \includegraphics[width=8cm]{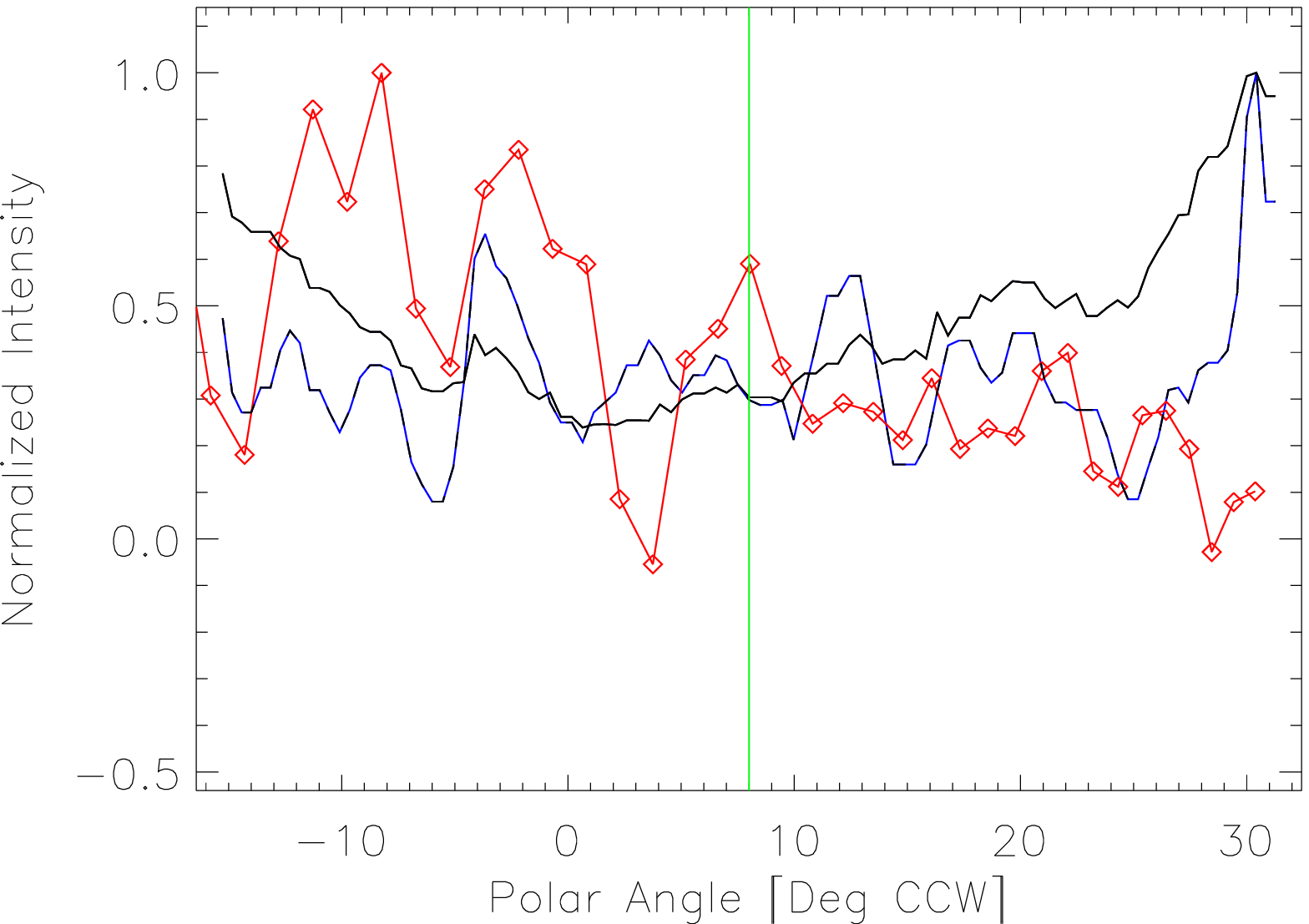}\\
  \end{tabular}
  \vspace{-0.2cm}
   \caption{Sequence of normalized intensity cuts at 2.42\RSUN, of C2 LASCO background subtracted images (black) 
  and wavelet-processed images (blue), with superimposed the normalized Si XII 499 \AA\ intensity observed along the UVCS slit (red).
  The top left panel shows the pre-CME LASCO C2 on 2001 May 12  at 22:15 
  	and UVCS Si XII 499 \AA\, at 1.97\RSUN\, from 21:34 to 22:25 UT.
  The top right panel shows the WL ray on 05/13 at 02:41 UT  
  	and UVCS Si XII 499 \AA\, at 2.42\RSUN\, from  03:48 to 05:31UT. 
  The bottom left panel shows the WL ray on 2001 May 13 at 05:50 UT  
  	and UVCS Si XII 499 \AA\, at 1.67\RSUN\, from 05:33 to 06:05 UT. 
  Finally the bottom right panel shows the post-ray WL corona on 2001 May 13 at 19:30 UT 
  	and UVCS Si XII 499 \AA\, at 1.63\RSUN\, from 19:28 to 20:17 UT.
  The intensities are the same as in Figure~\ref{c2_010512}. Green line marks the polar angle of the WL ray.}
   \label{pl_010512}
\end{figure*}

{\it Coronal Ray (2001 May 12)} - The LASCO C2 images of the events are in Figure~\ref{c2_010512} and show the 
pre-CME corona (top row), the ray in the wake of the CME (middle rows) and the post-ray corona (bottom row). 
The figure also shows the Si XII 499\AA\ images along the UVCS slit. Before the CME and after the ray disappeared 
the UVCS slit was located below the occulter disk while during the 
lifetime of the ray, UVCS observed at 1.67 and 2.42 \RSUN. The ray is clearly visible in the background 
subtracted and wavelet-processed images. Si XII 499\AA\ images show a narrow bright area at the position of the ray. 
Similar peaks are also observed in O VI and \LA lines indicating typical coronal temperatures. This narrow bright area is not
present in the UV spectra of the pre-CME and post-ray corona. As with the previous event the normalized intensities extracted 
along the white dashed lines 
in LASCO images or at the same location of the UVCS slit when it is located above the occulter and  \Si12 499 \AA\ are plotted in 
Figure~\ref{pl_010512}. In the top right panel WL and \Si12
intensities are at the same height, 2.42\RSUN, while in the bottom left panel Si XII intensity was detected at 1.67 \RSUN.
Again the WL ray and UV bright narrow spots are well aligned.

\begin{figure*}
\centering
\begin{tabular}{cc}
   \includegraphics[width=6.0cm]{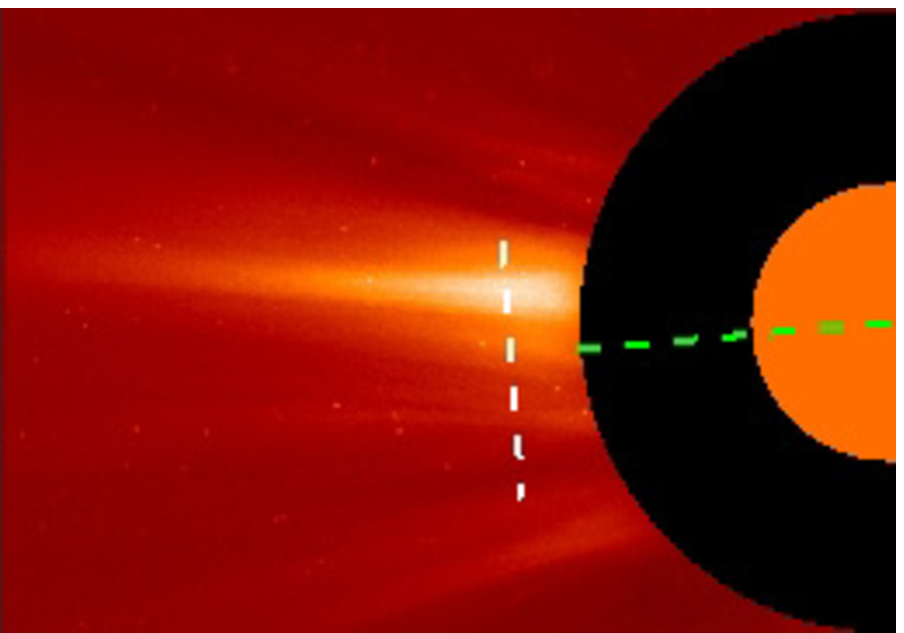} &	\includegraphics[width=6.0cm]{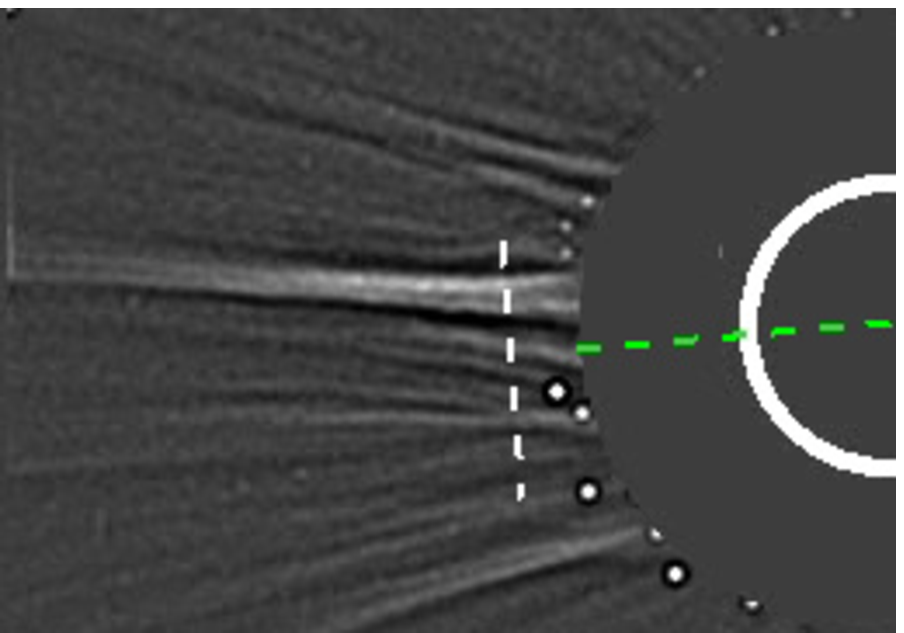}\\
   \includegraphics[width=6.0cm]{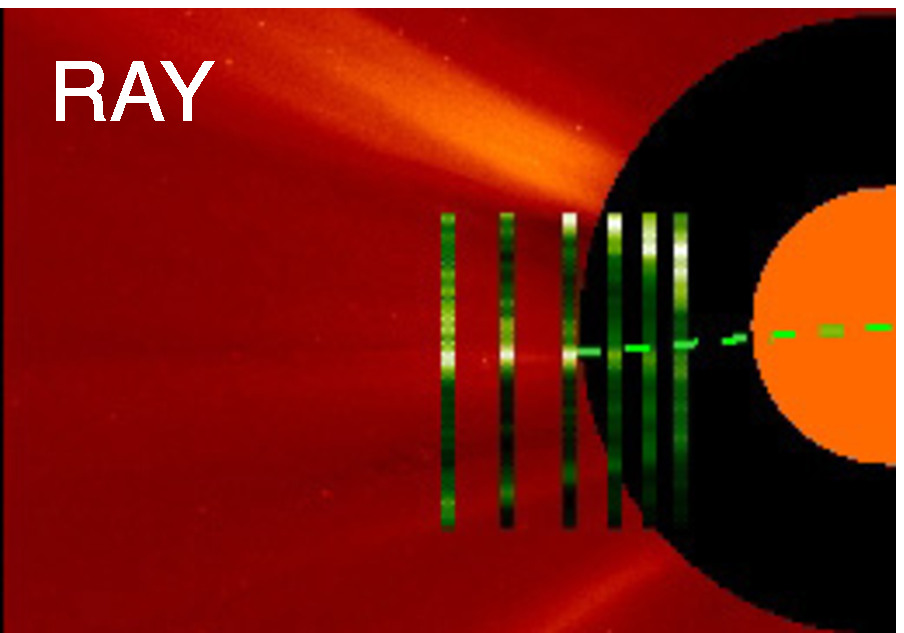} &	\includegraphics[width=6.0cm]{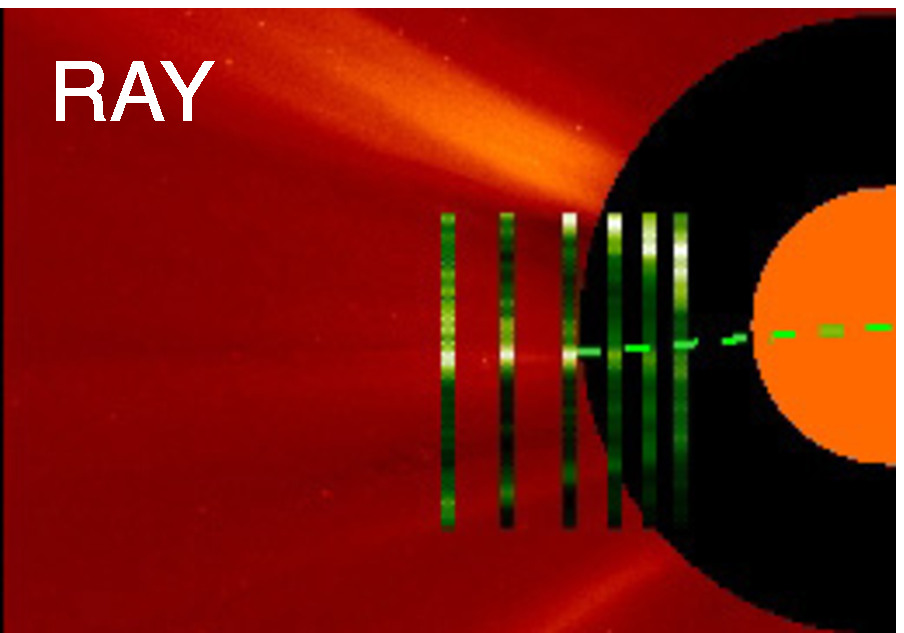}\\
   \includegraphics[width=6.0cm]{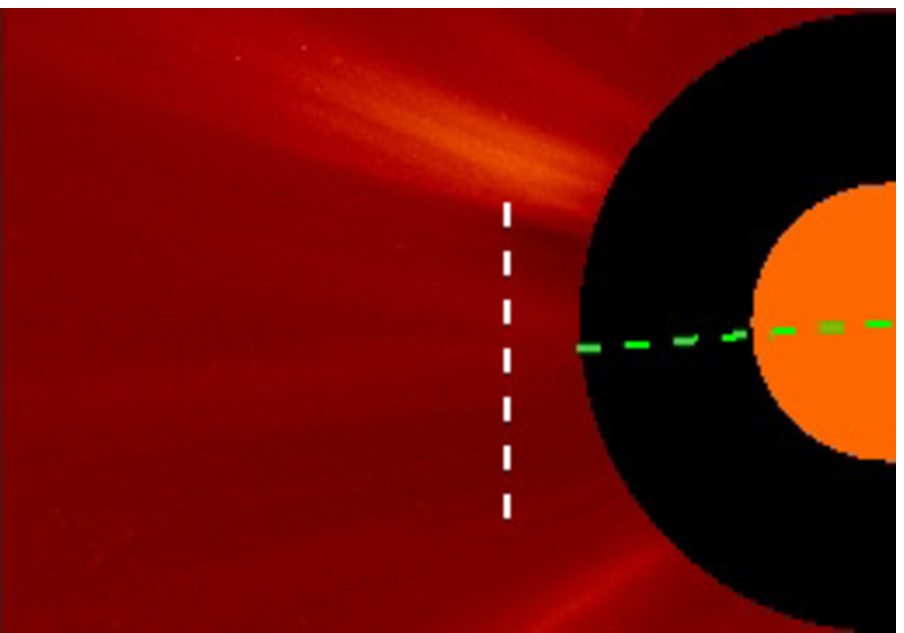} &	\includegraphics[width=6.0cm]{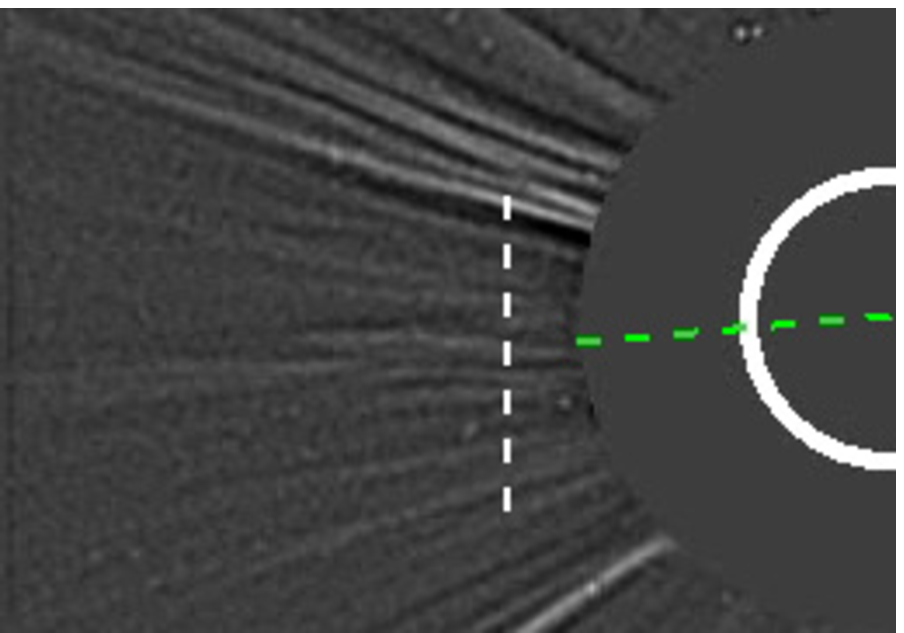}\\
\end{tabular}
   \caption{LASCO C2 images and wavelet-processed of the ray associated to the CME on 1998 
      April 23 at 05:27 UT. 
     The top panels show the pre-CME corona on 1998 April 23 at 03:55 UT. 
 The middle panel show the ray at 15:27 UT with 
     superimposed the \Osix 1032\AA\ intensity image at several heights along the UVCS slit.
     The bottom panel the post-ray corona at 22:27 UT. The  green dashed line marks the position of the white light ray, 
     while the white dashed line mark the position along which LASCO data have been extracted for Figure~\ref{pl_980423}.}
   \label{c2_980423}
\end{figure*}

\begin{figure}
\centering
  \includegraphics[width=9cm]{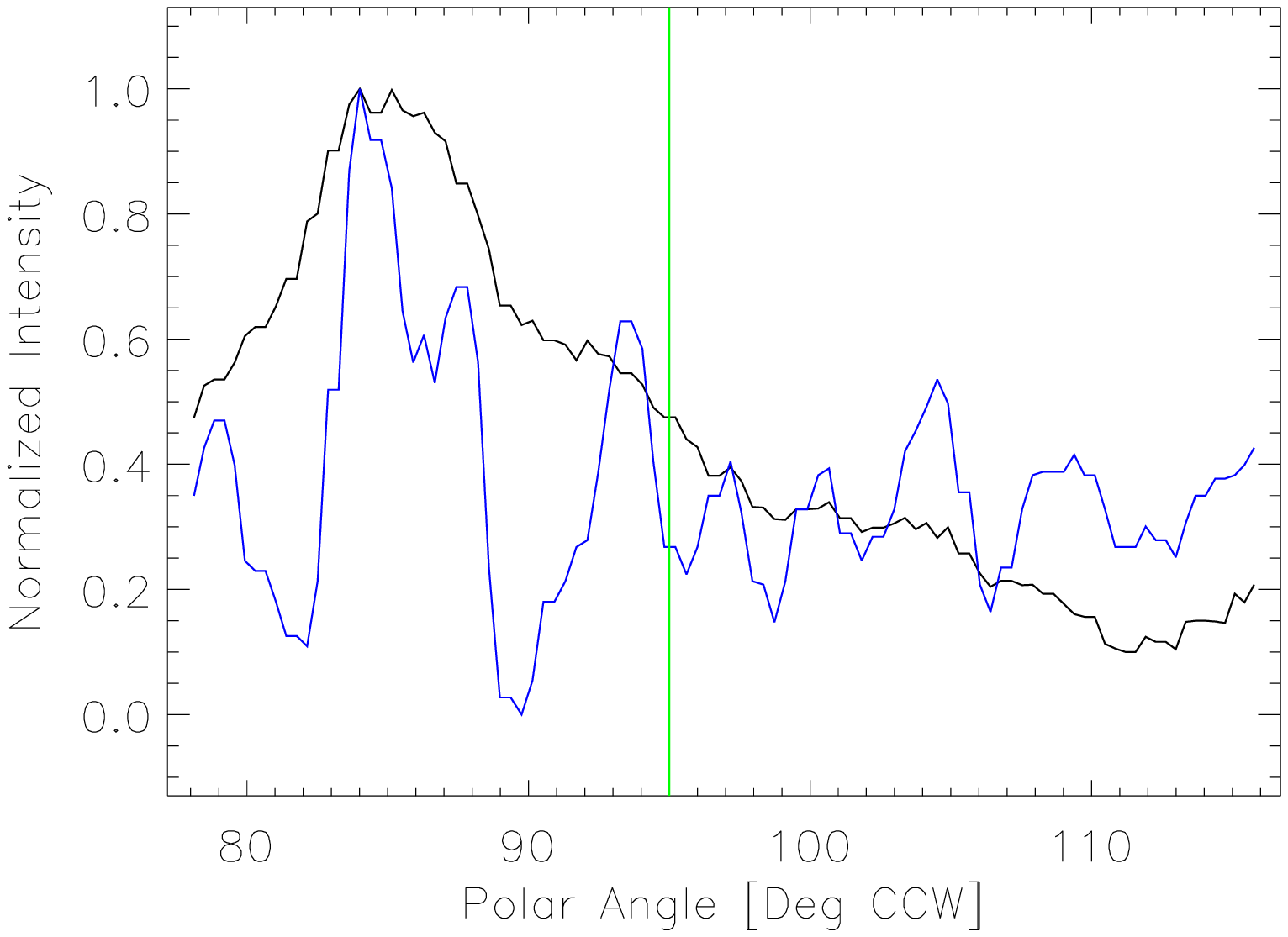} \\
  \includegraphics[width=9cm]{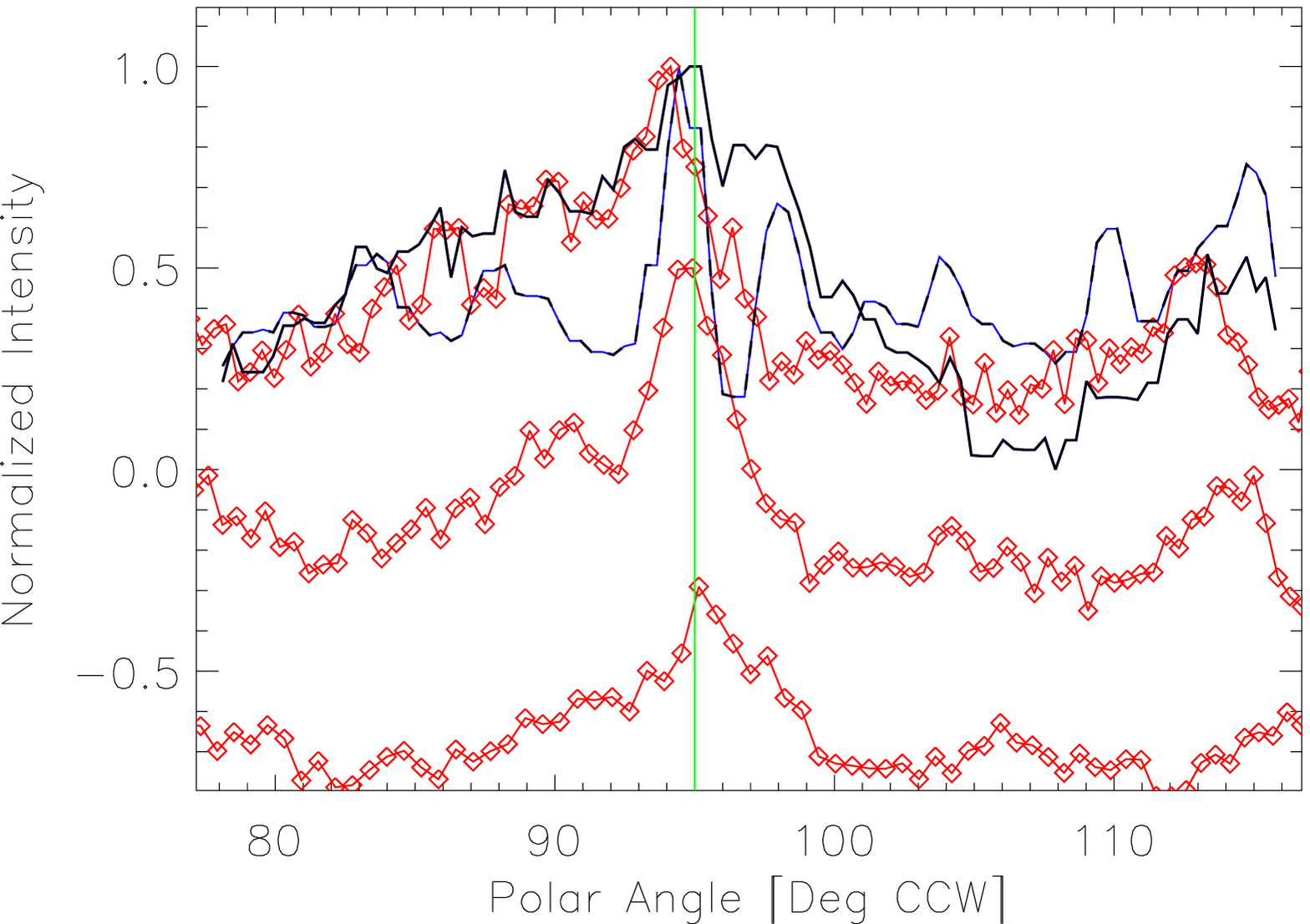} \\
  \includegraphics[width=9cm]{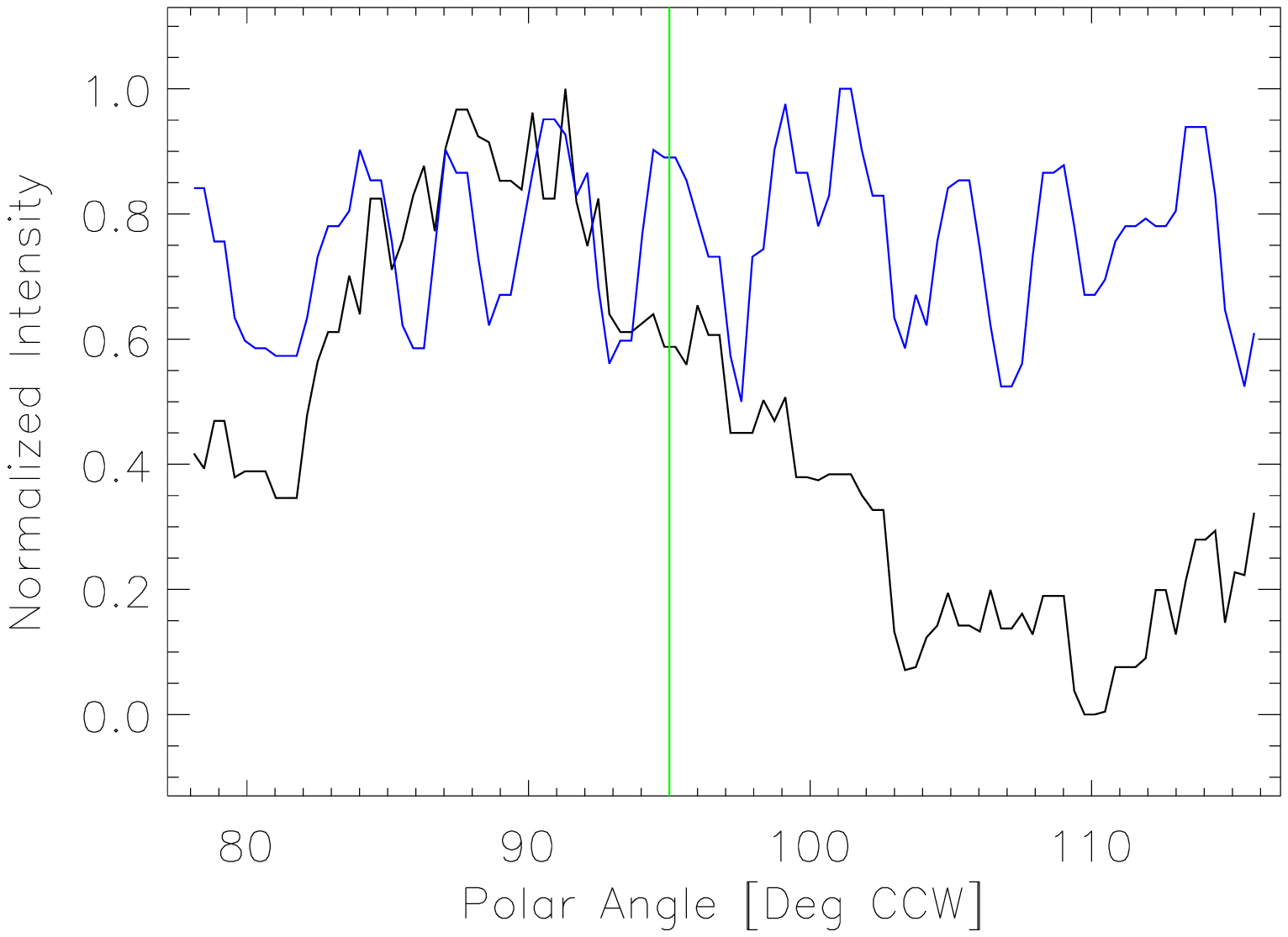} \\
   \caption{Sequence of normalized intensity cuts at 2.47\RSUN, of C2 LASCO background subtracted images (black) 
  and wavelet-processed images (blue). Green line marks the polar angle of the WL ray. 
  Upper panel shows LASCO C2 pre-CME on 1998 April 23 03:55 UT, 
  middle panel shows LASCO C2 ray on 1998 April 23 15:27 UT and \Osix 1032\AA\ at 3 different heights: 
  2.74\RSUN\, (upper red line) on 1998 April 23 from 14:46 to 15:04 UT,
  2.29\RSUN\, (middle red line) on 1998 April 23 from 14:27 to 14:45 UT, and
  1.95\RSUN\, (lower red line) on 1998 April 23 from 14:13 to 14:26 UT. The \Osix line intensities have been offset in order
   to show them in the same plot.
  Lower panel shows LASCO C2 post ray on 1998 April 23 22:27 UT.
  The intensities are the same as in Figure~\ref{c2_980423}.}
   \label{pl_980423}
\end{figure}

\begin{figure*}
	\centering
  \begin{tabular}{ccc}
	\includegraphics[width=5.1cm]{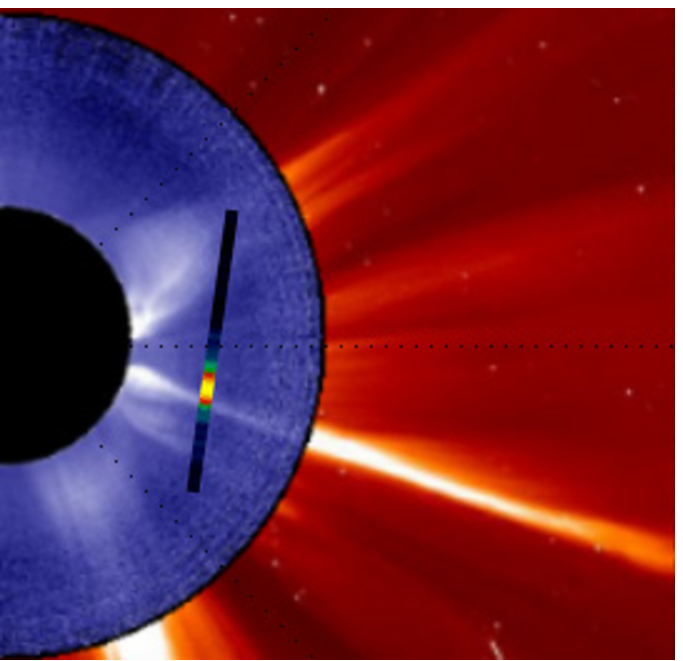} &
        \includegraphics[width=5.1cm]{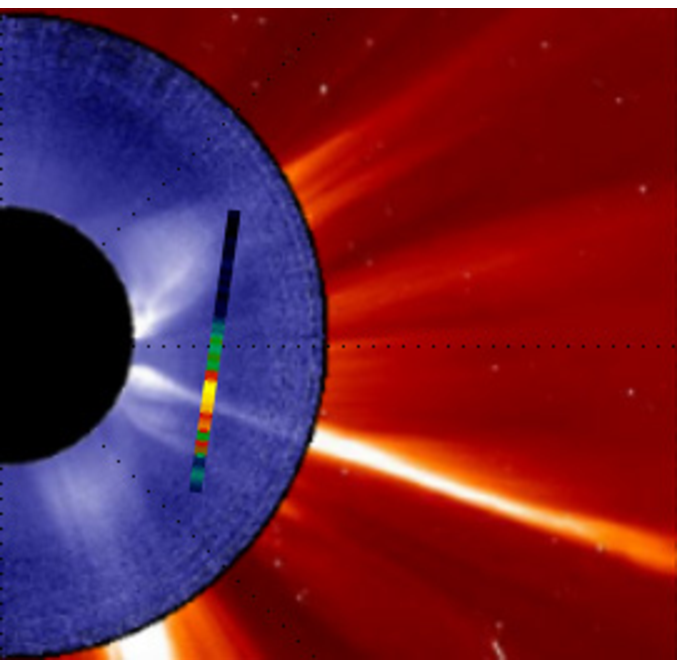} &
	\includegraphics[width=5.1cm]{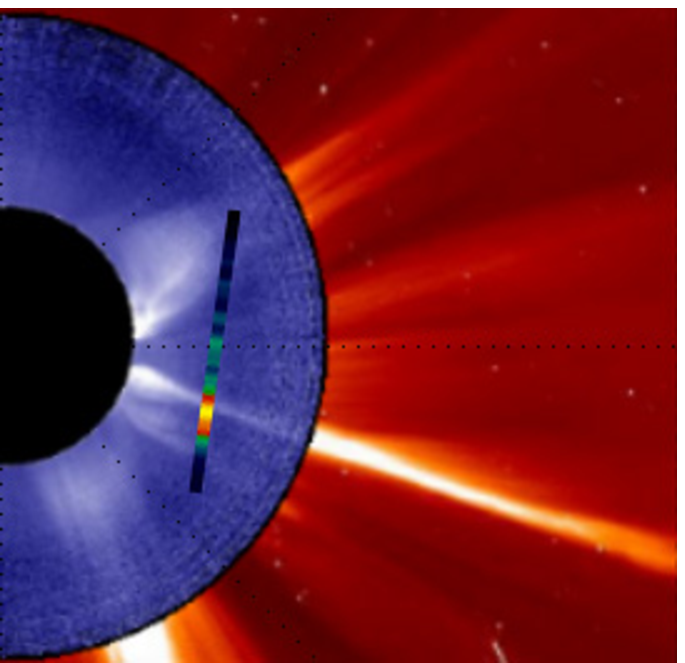}  \\  
  \end{tabular}
  \vspace{-0.2cm}
  \caption{Composite image of the ray detected on 2003 November 4 by LASCO C2 
  at 23:54 UT, MK IV at 23:57 and UVCS from 20:12 to 20:51 UT. The three 
  panels show UVCS images of \fe (left panel), \Si12 (middle panel) and 
  \Osix lines (right panel).}
\label{nov4_03}
\end{figure*}

\begin{figure}
\centering
   \includegraphics[width=9cm]{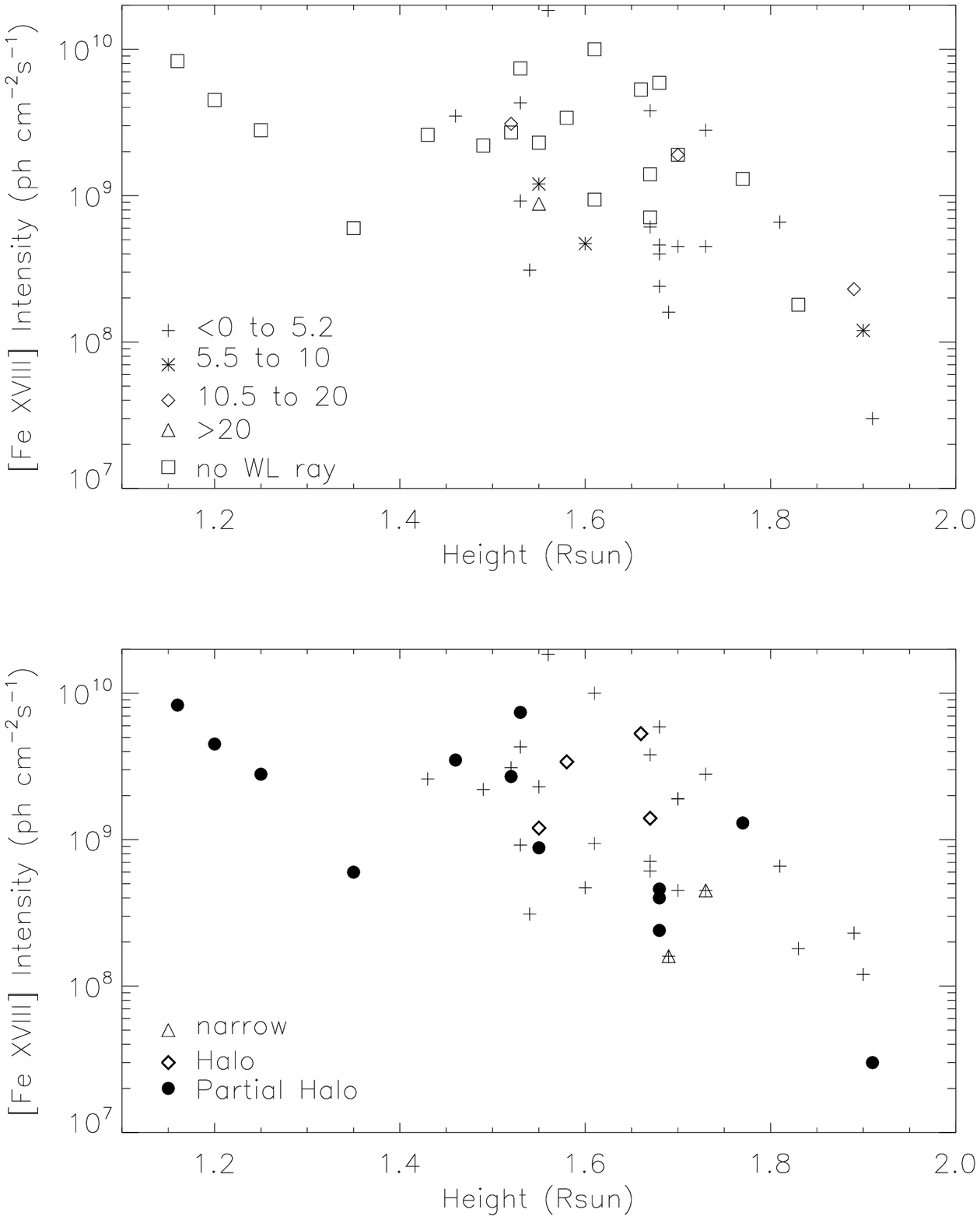}
   \caption{\fe line intensities as function of the heliocentric distance. The symbols refer to different delay time in hour (top panel),
     and CME morphology (bottom panel). In the bottom 4 CMEs are halos, 7 partial halo and 2 narrows with width of 14 and 44\DEG.}
   \label{fe_int}
\end{figure}

\begin{figure}
\centering
   \includegraphics[width=9cm]{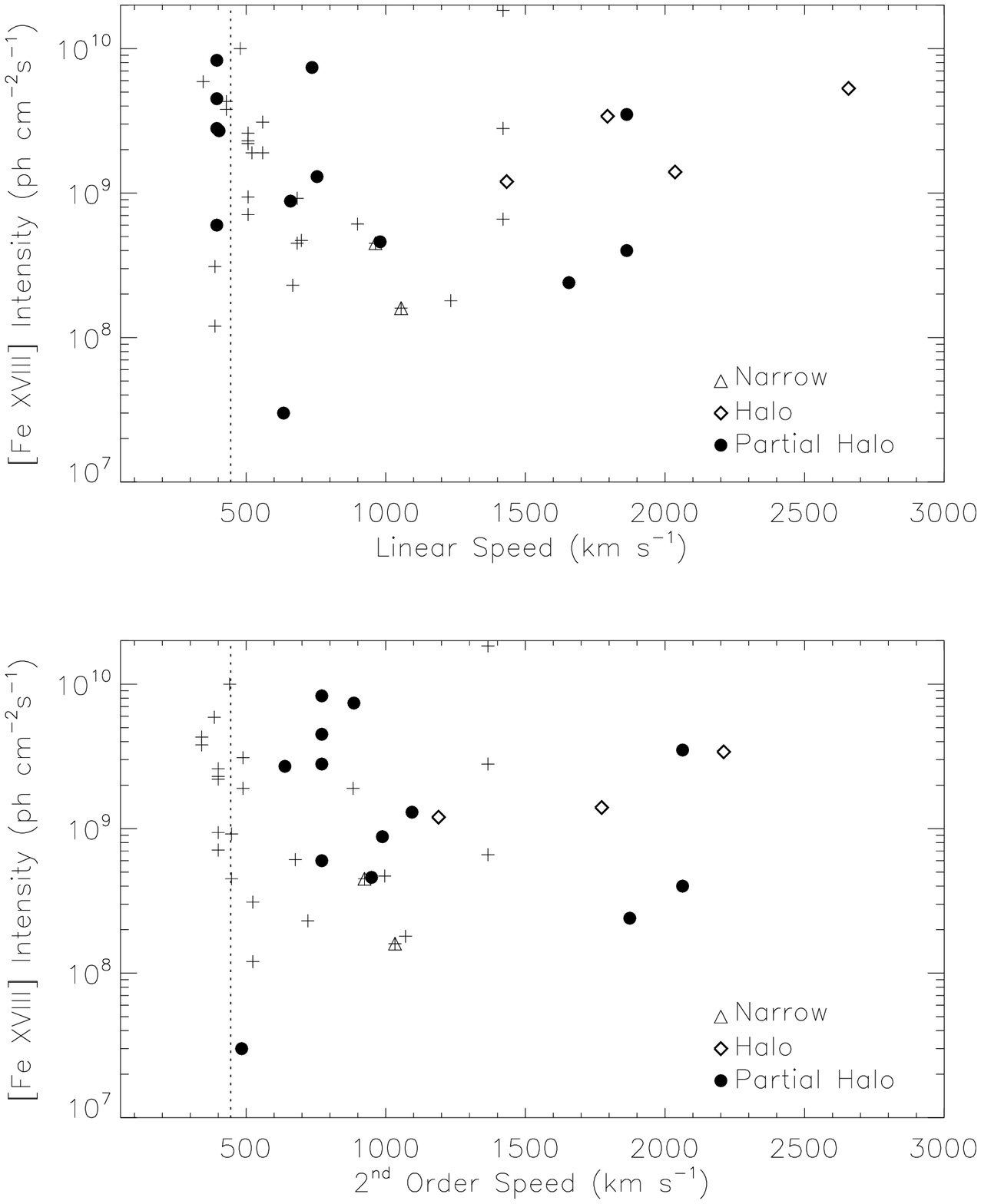}
   \caption{\fe line intensities as function of the CME linear (top panel) and second order speed (bottom panel). Triangles, diamonds and dots
     indicate the rays associated to narrow, halo and partial halo CMEs respectively. The vertical dotted line marks the average speed of CMEs during 
1996-2007 period.}
   \label{fe_speed}
\end{figure}

{\it Cool Ray (1998 April 23)} - A cool ray is that associated with the CME on 1998 April 23 at 05:27 UT.
The ray appeared in WL images at 10:55 UT at PA 95\DEG. LASCO background subtracted and wavelet-processed images are shown in 
Figure~\ref{c2_980423}. In UVCS the ray is observed in \LA and \Osix at several heights but not in \Si12. In the middle panel 
of Figure~\ref{c2_980423} the \Osix intensities show very clearly the ray at 2.74, 2.29 and 1.95 \RSUN. The
UV lines overlap very well with the WL ray as is  also shown in the normalized intensities of Figure~\ref{pl_980423}. 
In this figure the WL intensities were extracted along the white dashed line of Figure~\ref{c2_980423} at a height of 2.4 \RSUN
and compared in the middle panel with the UV \Osix 1032\AA\ intensities at three heights.

From the 60 of the 157 WL-selected rays which were also detected in UV spectra,
only 10 were observed in the \fe line.  Of the remaining 
rays, 34 did not have the necessary wavelength range to observe the \fe line, and in 16 
the \fe line was 
within the wavelength band but not observable due to blending with the bright \LA line in
the redundant path. 
For further analysis, we have reorganized the 60 WL selected rays and the 18 UVCS selected
features into a set of events that show \fe (Table~\ref{fe_ray}) and a set that do
not (Table~\ref{nofe_ray}).  The \fe provides a strong constraint on the plasma temperature,
and events where it is measured merit more detailed analysis than the events that
do not.

A common characteristic of the UV rays is that hot and coronal
lines are often offset. An example is shown in Figure~\ref{nov4_03}. 
The figure shows the ray observed on 2003 November 4 after the CME at 19:54 UT \citep{Cia08}. 
The comparison among the lines of \fe (left panel), 
\Si12 \L521 (middle panel) and \Osix \L1032 (right panel) shows the displacement between coronal and hot lines.
While characteristics of the rays such as line intensities and widths, are quite simple to compute in 
those lines not usually present in the solar corona, e.g. \fe and \ca, the separation of the ray peak 
from the surrounding background
corona is sometimes ambiguous for typical coronal lines, e.g. \Si12, \Osix, \LA, etc. 
For that reason we report the line intensities and widths for the rays observed in \fe, and in those cases 
we also report the intensities for the cooler lines.  We select the same spatial regions for all the
lines, subtracting a background based on the intensity distribution along the slit. The widths were computed as FWHM of the spatial distribution of the \fe line.

\subsection{Hot Rays}

The 28 rays with \fe emission are listed in Table~\ref{fe_ray}. Only 10 of the rays selected
from the white light ray catalog show \fe emission. 
The remaining 18 candidates come from the UVCS CME catalog and 15 of them are outside the time periods selected for the LASCO study. 
It is important to note that the PA of the WL and UV rays in Table~\ref{fe_ray} are different for several reasons.
First, the PAs of WL rays are measured above the occulter disk (i.e. $\geq$ 2.5 \RSUN) at the time of 
first detection, while UV rays are mostly observed below the occulter, often later that the WL rays. 
Moreover, the rays are often not radial and move during their lifetime.
The temperatures listed in Table~\ref{fe_ray} were computed 
using the ratio between \fe and \Si12 lines under the assumption of ionization equilibrium. 
For the events in which only \fe was observed the temperature is not easily constrained. We used 'hot' 
to indicate that the temperature in the ray is higher than in the case in which \Si12 was also present. 

In twelve cases the narrow \fe emission has either no associated WL ray or a ray which did not satisfy the current
sheet selection criteria. For those events no delay time is listed. For the remaining rays the delay time
varies from cases in which the rays were detected earlier than in WL (negative numbers) to the most extreme UV observation where the ray was detected 21 hours later.
In some cases the ray has been observed at several heights.    
All the heights at which \fe has been detected are below the LASCO C2 occulter disk, with the highest at 1.9 \RSUN.

Figure~\ref{fe_int} shows the \fe line intensities as a function of the heliocentric distance for all the rays listed 
in Table~\ref{fe_ray}. In the top panel crosses, stars, diamonds and triangles correspond to different ranges of delay time. 
The squares are \fe bright narrow features with no corresponding rays in WL images. For those cases it was not possible to evaluate the delay time.  The \fe line intensity ranges over 
three orders of magnitude or, (0.3 - 190) $\times$ 10$^{8}$ \flux. 
In the bottom panel the rays associated with halos, partial halos and narrow CMEs are plotted using diamonds, dots and triangles, respectively. 
The \fe line intensity in the ray decreases with the height but it does not show 
correlation with the delay time of the observation. 
Rays associated with energetic events such as halo and partial halo CMEs do not show the highest intensities in \fe line. 
The four rays associated with halos have intensities above 10$^9$ \flux but the partial halos are scattered over the entire range 
of the observed intensities.  In halo and partial halo events, the ray might be far from the plane of the sky, so that
the actual position observed is greater than the apparent height.  This likely accounts for
the lack of correlation.

The \fe line intensity as function of the associated CME speed is plotted in Figure~\ref{fe_speed}. One might expect that
CME speed is related to Alfv\'{e}n speed and that the magnetic energy available to heat the plasma is roughly proportional to
$V_A^2$.  In the top and bottom panels
of the figure the CME linear and second order speed from LASCO catalog are used.  Rays associated
with narrow, halo and partial halo events are plotted with triangles, diamonds and dots respectively. The vertical dotted line
indicates the LASCO average speed (445 \kms) for normal CMEs during the period 1999-2007 \citep{Mit09}. 
More than 78\% of the hot rays are associated with CMEs with speed higher than the average LASCO speed.  
Again no strong correlation is seen in these plots although rays associated with partial halos and halos show  some correlation with 
the speed of the CMEs. Above 1200 \kms  \fe line intensities increase with the CME speed.

One might expect that the densities within a CME would be larger in more massive events, and that the correspondingly higher
emission measures would yield higher intensities. 
No mass and energy estimates are available in the LASCO catalog for partial halo, halo and narrow CMEs. For all the other rays 
no strong correlation between \fe line intensity and energy of the associated CME has been observed.  It is possible that this lack
of strong correlation results from the variation of temperature among the different events, since the \fe intensity depends on both
emission measure and temperature.

The spatial widths of the CS in Table~\ref{fe_ray} vary in the range (3 - 35) $\times$ 10$^9$ cm.  While these widths are far
larger than the widths expected for resistive reconnection in the corona, they are in reasonable agreement with predictions
of turbulent reconnection \citep{laz99} or with the exhaust region of Petschek reconnection half a solar radius from the
diffusion region, especially if the current sheet is not viewed edge-on.  The widths of the \fe features are generally 
much smaller than the widths of associated features seen in Si XII, O VI or Ly$\alpha$.  We found no dependence of the width on 
the heliocentric height, the observation delay, or CME speed. 

The ray temperatures range between ($\sim$ 3 - 8) $\times$ 10$^6$ K. 
The ten events in which the ray has been observed in the \fe line only, indicated with 'hot' in Table~\ref{fe_ray}, 
are not associated with particularly energetic events.  
The estimated temperature range is the range where the \fe  ionization fraction is large enough that the line is observable.  Evidence for higher current sheet 
temperatures is found in the observations of [Fe XXI] and Fe XXIV \citep{in03} and Fe XXIII \citep{re11, cheng}.  Recent
Hinode/Extreme Ultraviolet Imaging Spectrometer observations of an event that shows Fe XV but no higher ionization states provide evidence for a current sheet cooler than
$3 \times 10^6$ K \citep{landi12}.

\subsection{Spatial Offsets}

Half of the rays listed in Table 2 show a clear offset between the spatial positions of
the peaks on \fe and Si XII.  Figure~\ref{nov4_03} is a good example, in which the [Fe XVIII] peaks
on the northern edge of the ray, Si XII is a broad feature centered on the ray,
and O VI peaks on the southern edge of the ray.  The current sheet studied by
\cite{Ko03} is another good example (see Figure 12 of that paper).  In a few cases, Si XII or O VI show
two peaks, with the \fe peak lying between them.  In the cases where there is no
offset, the \fe peak is usually narrow, while the Si XII peak is broader and
the O VI peak is still broader. 

One possible interpretation of the offset is that
the cooler material is ambient coronal plasma about to be swept into the current sheet.  The
transverse motions seen in the Extreme-ultraviolet Imaging Telescope images by \cite{yokoyama}, 
Ly$\alpha$ emission observed with UVCS by \cite{Lin05} or AIA images by \cite{takasao} show inflow speeds
of the order of 5 to 100 $\rm km~s^{-1}$.  At those speeds, the coronal material
would be swept into the current sheet on timescales of order 1-10 hours.  Since the coronal gas is
generally located to one side of the hot material, the inflow into the current sheet
would be very asymmetric, a situation simulated by \cite{murphy}.  This can lead to
transverse motion of the current sheet, as reported by \cite{Ko03}.

An alternate interpretation of the offsets is that the cool material is essentially separate
from the current sheet, occupying nearby field lines in a streamer-like structure, in which
the inflow time to the current sheet is longer than the flow time along the magnetic
channel.  This interpretation is suggested by the cool blobs that move beside the hot
current sheet in the 2003 November 4 event \citep{Cia08}.  

In general it is difficult to discriminate between the two interpretations above, but detailed
study of individual events should make it possible to distinguish between them in some cases.

\subsection{Projection Effects}

Since we are observing optically thin structures, it is difficult to tell whether the apparent
width we observe is the actual thickness of the current sheet or the result of thin sheet of
gas viewed at an angle.  In a few cases the projection effects can be shown to be small,
as in the observation of a gap (i.e., a region of very low line intensity), rather than a bright region, in Ly$\alpha$ \citep{Lin05} and
the case in which the column density in white light and the emission measure from UVCS can
be combined to separate the density and thickness of the structure \citep{Cia08}.  The offsets
discussed in the previous section show that projection effects cannot be too large in those
events, as the \fe and the other lines would be superposed if the angle to the line of sight
was substantial.

\subsection{Coronal Rays}

Table~\ref{nofe_ray} presents a list of events where \fe emission was not observable due to the UVCS instrument
configuration (indicated by -) or observable but not detected (indicated by x).  We also indicate whether
the ray was detected (indicated by +) or not in the \Si12, \Osix and \LA lines.  Of the 50 events in Table~\ref{nofe_ray},
35 show both \Si12 and \Osix lines, so their temperatures are at least as high as that of the ambient corona.
One shows \Si12 without \Osix or \LA, indicating a temperature substantially higher than
the ambient corona. However, as mentioned above there are often offsets between the features seen
in different lines, so in some cases there may be a high temperature adjacent to a cooler one.  For
example, the 2003 November 4 event shown in Figure~\ref{nov4_03} has a bright ray in \fe and \Si12, but 
\Osix and \C3 emission are offset several degrees to the north.
 
No correlation has been observed between the rays with coronal temperatures and the speed or energy of the associated CMEs.
The range of linear speed of those rays varies between 100 and 1500 \kms. In particular the ray with the highest temperature,
'hot' in Table~\ref{nofe_ray}, is associated with a CME of speed $\sim$ 100 \kms.

\subsection{Cool Rays}

Table~\ref{nofe_ray} also lists 14 events that were detected in \Osix and/or \LA, but not in \Si12 or \fe.
The ranges of delay time and height of UVCS observations are similar to those of the coronal or hot rays.
These features are cooler than the ambient corona.  In some cases, lines of \C3 or \Ofiv were detected,
indicating much lower temperatures, on the order of $10^5$ K.  In some of those events, the
line width of the \LA line was clearly much smaller than the width of the coronal \LA line, indicating
proton temperatures of $10^5$ K or less.  Bright, long-lasting ray-like features in cool emission
lines in the wakes of CMEs are not uncommon in UVCS spectra, but in most cases they do not satisfy
the criteria for current sheets in LASCO.  They are rather surprising, in that they often last 
for many hours, which is much longer than the free-fall time from heights of 1.5 to 2 \RSUN.  These
features may be prominence plasma from the CME core slowly draining back along particular magnetic field lines
from much larger heights. Downflow  of hot and cold material in the wake of a CME associated to an erupting prominence has been reported by \cite{Tri07}.
As for the coronal rays, the cool rays are associated with CMEs with a wide range of 
speed and energy, including one halo and two partial halos.

\subsection{No UV Ray}

Many of the WL selected rays, 97 out of 157, did not show any corresponding UVCS features, mostly because UVCS was observing a different region during the WL ray lifetime.  
We do not include them in this study, but they will be considered in a paper on the white 
light rays alone (Webb 2012, in preparation).
In a few cases, there may be no identifiable UV feature because the temperature was so high that 
it lay outside the $10^4$ to $10^7$ K range covered by the UVCS lines.  In
that case, it might show up as a gap in all the lines, as in the event discussed by \cite{Lin05}.  It might also appear in hard X-ray observations as in the event reported by
\cite{sui03}.  
More commonly, however, the UVCS data were not taken at the right position during the duration of the WL ray. 
In other cases UVCS data were too noisy due to short exposure time or confusion with
other CME features, or the height of observation was too large.

\section{Summary and Conclusions}\label{concl}

A sample of 157 WL rays observed by LASCO, with an outward moving U or V-shaped structure detected in the wake of CMEs, 
has been used to search archival UVCS data in order to study their UV spectra. 
The sample included rays selected during both solar minimum (July 1996 - December 1998) and solar maximum (year 2001).
In 60\% of the WL sample, no UV rays were seen. Aside from the cases in which the UVCS instrument was observing 
a different target during the lifetime of the WL rays, a temperature higher than 10$^7$ K or a modest plasma emission 
measure would make the ray too faint for UVCS to detect.
Sixty rays (40\%) of the WL sample showed UV emission.  Of those 60 WL rays, 14, or 23\%, showed only cool
gas, probably from prominence material draining back from the CME core.  Of the 26 events for which \fe could be
detected if present, 10 of them, or 38\%, did show [Fe XVIII].  Many of the others had ordinary coronal temperatures,
but a few showed strong enough Si XII emission to 
indicate relatively hot gas.  Other rays may have been too hot to show \fe (T$> 10^7$ K).  We conclude that 
about 23\% of the rays selected from LASCO images are magnetic flux
tubes containing cold material, about 18\% are hot features consistent with the current sheet interpretation, and
the remaining 59\% with ordinary coronal temperatures may be more quiescent streamers that have reformed after
the eruption.  

We also considered a sample of 18 more rays detected in the \fe line by UVCS. Most of them were outside the periods
selected from the LASCO study. Ten out of 18 UV selected rays had no associated WL ray. While it is difficult to
quantify the sensitivity of the white light search, since in many cases it is determined by the complex background
of the evolving post-CME structure, it may be that the cases with no white light ray are examples of localized  hot
features with relatively small density enhancements or thicknesses.
  
Thus, WL images and UV spectra were analyzed for a total of 78 rays. The CME characteristics of the sample are quite broad in 
terms of speed, angular width and energy. Eight halo and 12 partial halo CMEs are in the sample.  
70\% of the rays were associated with CMEs with linear speed higher than the average speed of 445 \kms obtained by \citealt{Mit09} for
1996-2007 period. This is expected as  about 64\% of the rays were detected during 2000-2003, the long lasting solar maximum of 
Cycle 23, and more than 50\% of the 28 rays detected during solar minimum have speed higher than 445 \kms. 

The 78 rays were divided in two sub-samples with 
(see Table~\ref{fe_ray}) and without (see Table~\ref{nofe_ray}) \fe emission. The high temperature line \fe is not usually present 
in the solar corona and the rays observed appear more distinct from the background in this line than in typical coronal 
lines such as \Si12, \Osix and \LA, providing a simpler estimate of the line intensity and width of the 
rays, and strong constraints on their plasma temperature. 
Based on the lines detected in the UV spectra the temperature of the rays were estimated. 
37\% of the rays had high temperature as indicated by the presence of \fe emission, 45\% had temperature similar 
to the surrounding corona as indicated by the presence of \Si12, \Osix and HI \LA lines, and 18\% were cool rays 
detectable only in \Osix and/or HI \LA lines with temperatures around 10$^5$ K.  
Of the WL rays without \fe, only in 14 cases was the line in the observed wavelength range and not detected. 
In all the other cases especially in those that had \Si12 we cannot exclude that \fe could have been present as well.

Rays with \fe emission are mostly (85\%) associated with CMEs occurring during solar maximum. 
In the sample of rays with no \fe only 52\% are during solar maximum. 
Although the range of CME speed associated with the rays is very wide in both samples, the average and 
median speeds of the rays that show \fe are higher than those that do not. 
All the rays showing \fe emission have been observed at heights below the LASCO occulter disk.  

The temperatures of the rays detected in \fe line are higher than 10$^{6.5}$ K, but this is essentially a selection
effect due to the formation temperature of \fe.  No dependence on the height has been observed. 
The spatial widths range between 3 and 35 $\times$ 10$^9$ cm and again no correlation 
with the speed, delay and height exists.
The WL widths of four of the hot rays have also been estimated by \cite{song12} at 
3 and 4 \RSUN, and they are in agreement with those obtained from the \fe emission. 
There is a trend of \fe intensity to decrease with the height. This is also observed in the few cases 
in which the same ray was observed at different heights. 

A wide range of \fe line intensities has been detected, but no strong correlation has been observed with the 
delay time of the UV observations, or morphology of the associated CMEs.  The intensity of the \fe line in the 
ray associated with the 2002 January 8 CME \citep{Ko03}, measured with about 50 h delay, is among the highest.
The four rays associated with halo CMEs have \fe intensities well within the range of 
intensities detected for the CS associated to other CMEs. 

No significant correlation of the \fe intensity has been observed with CME speed or energy except for 
the halo and partial halo events. Above 1200 \kms there is a weak trend of \fe intensity to increase with the speed. 

The results outlined above provide a wide range of UV properties of WL rays, but no strong
correlations with the associated CME speed, energy and morphology has been found. 
While the hot and coronal rays may fit within the framework of the reconnection processes 
that bring ambient plasma into CS structures, the much lower temperatures observed in the 
cold rays suggest a different interpretation. These may be channels of cold prominence 
material draining back from the CME core as it moves outwards through the solar corona.

\acknowledgements
This work was performed under NASA grants NNX09AB17G-R and NNX11AB61G to the Smithsonian 
Astrophysical Observatory.  It was initiated during a series of workshops on {\it Understanding the role of current 
sheets in solar eruptive phenomena} at the International Space Sciences Institute (ISSI). We also benefited from 
the SOHO/LASCO CME catalog, generated and maintained by the Center for Solar Physics and Space Weather, The 
Catholic University of America in cooperation with NRL and NASA.

\clearpage

{\tiny{
\begin{deluxetable}{l l c c c c c r c c c c c c c l }
\tabletypesize{\scriptsize}
\tablecaption{Rays with \fe Emission \label{fe_ray}}
\tablewidth{0pt}
\tablehead{ 
CME 	& \multicolumn{3}{c}{LASCO Ray} 	& MK4 	& Sel 	& $\Delta t$ 	&\multicolumn{3}{c}{UVCS} & \multicolumn{4}{c}{Lines} 	                 	& Width		             & Log T 	 \\
 		& \multicolumn{3}{c}{} 			&  		& 		& C2/	&\multicolumn{3}{c}{}             & \multicolumn{4}{c}{10$^8$ \inte} 	& $10^9$ cm	  &             	 \\
		& Time	& PA 	& Durat. 		&  		&       	& UVCS		&Time      & PA   & \RSUN     &Fe XVIII & Si XII & OVI  &Ly$\alpha$  &                      &             }
\startdata
97-05-05 06:30$^a$ 	&-  			& -   			&   -		&  		  & W	  &     	& 05:16:38  & 71 		& 1.35	  & 6.    & 157  	& x    & x  	& 7  	&   		\\
               				&       		&     			&		&  	  &  	&     	&05:16:59 &69 		& 1.25	& 28   & 228  	& x    & x	& 6  	&   		\\
               				&       		&     			&		&  	  &  	&     	&05:17:21 &69 		& 1.20	& 45   & 420  	& x    & x	& 3  	&  		\\
               				&       		&     			&		&  	  &  	&     	&05:17:36 &69 		& 1.15	& 83   & 900  	& x    & x	&  4 	& 6.5  	\\
98-03-23 09:33$^b$		& -  			& -   			&  -		&N 	& U	& -   	&23.16:00 &257	& 1.52	& 27   & 78   	& x    & x	& 12 & 6.7 	\\
98-04-20 10:07$^c$		& 15:27 		& 262-250	& 14		&-   	& W	& 4.0	&20.19:30 &247	& 1.46	& 35   & x    	& x    & x	&    	&  hot	\\    
               				&       		&      			&		&   &  & 4.2 &20.19:36 &247& 1.68& 4.0   & x    & x    & x    & 6  &  "\\ 
00-02-22 14:06 		&  -    		& -    			& -		&   & U&     &22.18:33 &47 & 1.43  & 26  & x    & x    & x    & 9  & hot   \\
               				&       		&      			&		&   &  &     &   18:50 &48 & 1.49  & 22  & x    & x    & x    & 12 & "   \\
               				&       		&      			&		&   &  &     &   19:01 &45 & 1.55  & 23  & x    & x    & x    & 18 & "   \\
               				&       		&      			&		&   &  &     &   19:12 &46 & 1.61  & 9.4 & x    & x    & x    & 12 & "   \\
               				&       		&      			&		&   &  &     &   19:23 &45 & 1.67  & 7.1 & x    & x    & x    & 12 & "   \\
00-09-23 21:26 		&01:27/24	& 300 		& 21		&Y?&U& -3.1&23.22:19 &302& 1.67  & 38  & x    & x    & x    & 6   & hot         \\ 
               				&       		&      			&		& &  & -3.0&   22:29 &302& 1.53  & 43  & x    & x    & x    & 8   &  "      \\
01-01-26 16:06 		& 21:08 		& 52   		& ?		&Y?&W& 7.0 &27.04:06 &52 & 1.60  & 4.7 & 120  & 71   & 5.2  & 6   & 6.50   \\
01-07-05 05:54 		& 07:54 		& 127  		& 9		&-& W& 3.2 &05.11:07 &125& 3.12  & x   & x    & 0.5  & 5.5  &     & \\
               				&       		&      			&		& &  & 3.5 &   11:25 &126& 2.58  & x   & x    & 1.9  & 4.6  &     & \\
               				&       		&      			&		& &  & 4.0 &   11:52 &126& 2.13  & x   & x    & 1.7  & 24   &     & \\
               				&       		&      			&		& &  & 4.5 &   12:23 &124& 1.73  & 4.5 & x    & 39   & -    & 15  &  \\
               				&       		&      			&		& &  & 4.7 &   12:33 &130& 1.55  & x   & 10   & 130  & 720  &     &  \\
01-07-20 13:31 		& 18:06 		& 80   		&$>$9	&Y?& W& 15.4&21.09:27 &84& 3.11  & x   & x    & 0.95 & 350  &     &    \\ 
               				&       		&      			&        	& &  & 15.6&21.09:44 &84 & 2.56  & x   & x    & 3.2  & 110  &     &   \\
               				&       		&      			&        	& &  & 16.1&21.10:12 &83 & 2.12  &  x  & x    & 15   & 420  &     &    \\  
               				&       		&      			&        	& &  & 16.4&21.10:29 &81 & 1.89  & 2.3 & 380  & 12   & 5700 & 8   &6.42     \\  
               				&       		&      			&        	& &  & 16.6&21.10:42 &83 & 1.70  &  x  & 10   & 51   & 1300 &     &   \\  
               				&       		&      			&        	& &  & 16.8&21.10:52 &84 & 1.53  &  x  & 16   & 130  & 1900 &     &    \\  
01-08-25 16:50 		&17:50   		& 128 		& ?      	&Y& W& 6.7 &26.00:29 &125& 1.55  & 12  & 45   & x    & x    & 17  & 6.65   \\
01-09-21 08:54 		& 11:30  		& 117 		& 33.6   	&Y& W& 21.4&22.08:32 &115& 1.55  & 8.8 & x    & x    & x    & 13  & hot    \\
01-10-05 21:08 		& 21:30  		& 260 		& 21     	&Y& W& 4.8 &06.02:16 &263& 1.70  & 4.5 & 110  & 200  & x    & 8   & 6.50    \\
               				&        		&     			&        	& &  & 5.0 &   02:29 &261& 1.53  & 9.2 & 390  & 480  & x    & 6   & 6.52    \\
01-10-30 18:16$^h$ 	&    -   		&   - 			& -  		&-& U&  -  &31.04:27 &92 & 1.53  & 74  &   x  &   x  &  x   & 20  & hot  \\
01-11-01 03:30 		&09:54   		& 272 		&$\sim$38&Y?&W& 16.3&02.02:11 &273& 1.70  & 19  & 61   & 57   & 300  & 8   & 6.63    \\
               				&        		&     			&        	&  & & 16.5&   02:23 &273& 1.52  & 31  & 200  & 900  & 1000 & 6   & 6.58    \\ 
01-12-20 21:12 		&00:54$/$21	& 282 		&$>$8	&?& W& -1.6&20.23:20 &280& 1.54  & 3.1 & x    & x    & -    & 15  & hot     \\ 
               				&        		&     			&        	& &  & 8.4 &21.09:20 &278& 1.90  & 1.2 & 1.6  & x    & -    & 13  & 6.72     \\
01-12-29 09:54 		& 12:30  		&280  		& 3      	&-& U& -1.8&29.10:41 &282& 1.91  & 0.3 & x    &  x   &  x   & 6   & hot  \\    
02-01-08 17:54$^d$ 	& -  			& -   			& -      	&-& U& - &10.20:45 &83 & 1.58  & 34  & 420  & 30   & 1700 & 7   & 6.55\\
02-11-26 17:06$^e$ 	&  -   			&  -  			&  -   		&-& U& -    &27.02:00 &305& 1.61 & 100 &  x   & x    & x    & 6   & hot\\ 
03-01-03 11:30     		& -     		& -   			& -   		&N& U& - &03.11:35 &299  &1.70   & 19  &6.1   & x    & x    & 4   & 6.9  \\
03-06-02 00:30$^f$ 		& 01:54  		& 248 		&$>$5	&?& U&4.2&02.06:08 &248  & 1.68  & 2.4 &$<$0.4& 33   & x    & 27  & 6.9  \\
03-06-02 08:54$^f$ 		& 10:06  		& 260 		&2   		&Y?&U& 0 &02.10:06 &258  & 1.68  & 4.6 & 11   & x    & x    & 9   & 6.7  \\ 
03-06-13 16:30     		&    -   		& -   			&-   		&N &U& - &13.17:23 &280  & 1.68  & 59  & 190  & x    & x    & 8   & 6.65   \\
03-10-24 02:54 		& 05:54  		& 100		 & 11		&N &U& 1.6 &24.04:29&104 & 1.69  & 1.6 &  x   &  x   & -    & 18  & hot\\
03-10-24 05:30 		& -      		& -   			&  -     	&  &U&     &24.06:35  &128& 1.83  & 1.8 &  x   &  x   & -    & 6   & hot \\
03-11-01 23:06 		& 03:54$/$2	&250  		& 5.5  	&Y?&U&-3.8 &02.00:06 &249& 1.67  & 6.1 & 110  &  x   &  -   &35   &$>$6.65   \\
03-11-02 09:30 		&$\dag$  		&$\dag$		& $\dag$	&Y?& U& ? &02.17:01 &240 & 1.67  & 14  & 120  & x    & -    &7    & $>$6.75  \\ 
03-11-03 10:06 		& 10:55  		& 275-272  	&$\sim$5.6&N& U& 0.2 &03.11:07&267&1.81    & 6.6 & 30   & x    & x    & 20  & 6.75       \\
               				&        		&      			&       	& &  & 2.4 &   13:18&271&1.73    & 28  & 26   & x    & x    & 20  &     \\
               				&        		&      			&       	& &  & 2.6 &   13:31&271&1.56    & 184 & 128  & x    & x    & 20  &     \\
03-11-04 19:54$^g$		& -   			&-     			& -     	& & U&     &04.20:30 &253& 1.66  & 53  & 17   & x    & x    &11   & 6.90  \\
04-07-28 03:30		&  - 		& -   			& -		&N& U& - &28.03:47 &280  & 1.77  & 13  & 20   & x    & x    &15   & 6.70    \\
\enddata

\end{deluxetable}


Starting from the left, the table lists the CME onset time as given in  the LASCO catalog, the time at which 
the ray was first seen, the PA and the duration of the WL ray selected in LASCO, the MLSO/Mk4 data, and the 
observation used for selecting the ray, U for UVCS and W for WL. For the Mark4 we used '-' when the data were
not available, 'N' and 'Y' when the data were available and the ray was not observed or observed, respectively. 
A question mark indicates an uncertain detection or case where the ray was not clear. 
In the seventh column is the delay time, in hours, between the first observation in LASCO and in UVCS.
The eighth, ninth and tenth columns show the time, PA and height at which the ray was observed by UVCS.
From column 11 to 14 are listed the line intensities of \fe, \Si12, \Osix and \LA.  In these columns
'-' indicates that the line was not observable due to the instrument configuration while 'x' indicates that
the line was observable but not detected.
The spatial width of the \fe emission and the estimated temperature of the ray are in the last two columns. \\
$^a$  The UV ray is associated to the same CME as the ray selected from WL but it does not satisfies the selection criteria\\  
$^b$  Ciaravella et al. 2002. Many bright rays are seen in WL but not obiously related to the UV ray. \\
$^c$  The ray changes position during the observation\\
$^d$  Ko et al. 2003 \\
$^e$  Bemporad et al. 2006 \\
$^f$ Schettino et al. 2010 \\
$^g$ Ciaravella \& Raymond 2008 \\
$^h$ CDAW catalog incorrectly lists at 20:42:06 \\
$\dag$ many rays, not clear\\

\begin{deluxetable}{l r r r r c c c c c c c c}
\tabletypesize{\scriptsize}
\tablecaption{Rays with no \fe Emission\label{nofe_ray}}
\tablewidth{0pt}
\tablehead{ 
CME 	& \multicolumn{3}{c}{LASCO Ray}	& MK4	&$\Delta t$	&UVCS	&		& \multicolumn{4}{c}{Lines} 			& Log T \\
 		& Time 	& PA 	& Durat. 		& 		&C2/UVCS 	& Time 	&  \RSUN & Fe XVIII & Si XII & OVI &Ly$\alpha$ 	& }
\startdata
96-08-22 08:38 & 09:38  & 119 & $>$15h&N& 6.0 &22.15:39  &$\le$2.02    & - & x & + & + & cool \\
96-10-19 02:39 &3:56/20 & 126 & 20h   &Y& 16.2 &20.10:46  & 2.31- 1.45 & - & + & + & + & coronal \\
97-02-23 02:55 & 06:30 & 78   & ?     &-&  0.7 & 23.07:11 & 1.45-2.3   & - & + & + & + &  "   \\
97-04-22 22:57 &10:50/23 & 70 & 2.5d  &?&  2.8 & 23.13:04 &1.42-3.11   & - & + & + & + &  "   \\
97-05-07 10:26 & 22:16  & 277 & 18h   &-& 9.7 & 08.02:29 & 1.4-3.1     & - & + & + & x &  "   \\
97-07-22 08:57 & 11:02  & 250 & 1.5d? &-& 4.9 & 22.13:50 & 1.5-2.3     & - & + & + & + &  "   \\
97-07-25 21:01 &22:00   & 268 & ~1.5d &?& 4.0 & 26.01:48 & 1.42-4.1    & - & + & + & + &  "   \\
97-08-26 00:00 &10:53/27& 80  & 1d    &?& 18.1 & 28.08:57 & 1.46-3.15  & - & + & + & + &  "   \\ 
97-08-28 16:41 &08:0/29 & 96  & 1.8d  &Y?& -0.2 & 29.07:46 & 1.46-3.15 & x & + & + & + &  "   \\
97-09-13 11:33 &18:19/14& 90  &$>$1d  &-& 14.1 & 15.08:25 & 1.5-3.1    & - & + & + & + &  "    \\
97-09-19 01:29 &15:04  & 275  &$>$19h &-& 4.9 & 20.00:25 & 1.45-3.15?  & - & + & + & + &  "  \\
97-09-28 14:28 & 20:59 & 90   &~1d    &Y?& 12.9 & 29.09:56 & 1.5-3.15  & - & + & + & + &  "   \\
97-10-12 06:26 & 08:28 & 255  & 13h   &Y?& 4.6  & 13.02:04 & 1.5-2.3   & - & + & + & x &  "   \\
97-11-22 05:45 & 21:30 & 45   & 1d    &Y?& 1.5  & 23.03:31 & 1.5-2.3   & - & + & + & + &  "   \\
97-12-02 16:43 &12:27/3& 80   & 18h   &-& 15.6 & 04.08:59 & 1.5-2.3    & - & + & + & ? &  "   \\  
98-03-23 00:50 & 08:13 &  85  & days  &Y?&  1.1 & 23.09:06 & 1.5-3.17  & - & + & + & + &  "   \\
98-03-27 20:07 &01:04/28& 195 & 15h   &-& 10.8 & 28.11:55 & 1.5-2.07?  & - & + & + & + &  "   \\
98-03-29 03:48 & 10:06 &190   & $>$5  &-& 2.0  & 29.12:07 & 1.5-2.64   & - & + & + & + &  "   \\
98-03-31 06:12 & 08:20 &174   & $>$7h &-& 4.4  & 31.12:46 & 1.5-2.5    & - & + & + & x &  "   \\
98-04-21 19:03 & 23:27 & 280  & 1.7d  &Y?& -1.2 & 21.22:13 & 1.4-3.73  & x & + & + & + &  "   \\
98-04-23 05:27 & 10:55 & 95   & 7.5h  &N&  1.6 & 23.13:34 &1.48-3.17   & - & x & + & + & cool  \\
98-05-21 05:27 & 11:02 & 293  &17.5h  &Y& 15.5& 22.02:35 & 1.5 -3.2    & - & x & x & + &  "    \\
98-06-02 08:08 & 17:27 & 246  & 3h    &-& 0.1 & 02.17:30 & 2.3         & - & x & + & x &  "    \\
98-06-13 02:27 & 11:27 & 268  & 23h   &Y?& 3.0 & 14.00:30 & 1.48-1.95  & - & + & + & + & coronal \\
01-01-02 21:30 &11:54/3& 40   & 12h   &-& 16.7 & 04.04:47 & 3.6-1.52   & x & + & + & + &  "   \\
01-01-25 15:30 &03:06/26& 180  & 9h   &N& 2.0 & 26.05:04 & 4.15-3.6    & x & x & + & + & cool   \\
01-01-31 18:06 &10:54/02 & 55 &$>$9h  &?& 5.5 & 02.16:22 & 1.70-1.55   & - & + & + & + & coronal \\
01-02-19 19:50 & 23:00 & 270  & 20h   &?& 3.9 & 19.02:55 & 1.71-1.53   & - & + & + & + &  "   \\
01-03-04 05:50 & 08:26 & 40   & ?    &Y& 6.4 & 04.14:49 & 1.70-1.53   & - & x & + & + & cool    \\
01-03-18 02:26 & 06:06 &108   &  9h   &N& 10.6& 18.16:45 & 1.7-1.5     & - & x & + & + &  "   \\
01-03-26 00:26 & 13:50 &288   & 18.5h &N& 2.4 & 26.16:12 & 1.7-3.1     & - & + & + & + & coronal  \\
01-04-22 05:26 & 13:06 &160   & ?    &?& -0.1& 22.12:59 & 3.13-1.56   & - & + & + & + &  "    \\
01-04-30 14:30 & 16:54 & 88   & ?    &-& -0.1& 30.16:50 & 1.7-1.5     & - & + & + & + &  "   \\
01-05-12 23:00 &02:21$/$14&  8   &$\sim$13.5h&-&-0.1 & 13.02:14 & 2.47-1.74   & - & + & + & + &  "    \\
01-05-13$/$14 14:15 & 17:20&352    & 12h &-& 2.0 & 13.19:17 & 1.63-2.42   & - & + & + & + &  "     \\
01-05-31 03:08 & 07:31 & 247  & 24    &-& 12.2& 01.05:44 & 1.70-1.53   & x & x & + & + & cool   \\
01-06-02 11:54  & 13:54 &290   & 13h  &?& 2.6 & 02.16:33 & 3.1         & x & x & x & + &  "    \\
01-07-19 10:30  & 11:06 & 230  & 14h  &?& 5.8 & 19.16:54 & 3.1-1.53    & x & x & + & + &  "           \\
01-07-20 13:31  & 18:06 & 80   &$>$9h &Y?& 15.3& 21.09:26 & 3.11-1.53  & x & + & + & + & coronal    \\  
01-07-20 22:22  & 23:15 & 110  &$>$12h&-& 10.2& 21.09:26 & 3.11-1.53   & x & + & + & + &  "     \\
01-08-09 21:30 &02:06/10& 106  & ?   &?& 9.2 & 10.11:20 & 3.1-1.5     & x & x & + & + & cool   \\
01-08-21 12:06 &2:27/22 & 234  &$>$21h&Y?& 5.4 & 22.07:51 & 1.7-1.5    & x & x & + & + &  "    \\
01-10-01 05:30 & 07:31  & 200  & 24h  &?& 7.2 & 01.14:46 & 3.1-1.53    & x & + & + & + & coronal   \\
01-10-01 22:06 &15:30/2 & 40   & 48h  &Y&12.9 & 03.04:24 & 1.7-1.5     & x & x & + & + & cool   \\
01-10-13 05:54 & 18:39  & 130  & 3d   &?&23.3 & 14.15:59 & 1.7-1.5     & x & + & + & + & coronal       \\
01-10-16 09:50 &16:06/17& 323  & 1d   &-&18.9 & 18.10:59 & 2.1-1.5     & x & + & x & x & hot       \\
01-11-12 10:06 & 23:06  & 300  & 12h  &-& 3.9 & 13.03:00 & 3.1-1.5     & x & + & + & + & coronal       \\
01-12-12 18:30 & 23:54  & 113  & 24h  &-& 2.5 & 13.02:26 & 1.5-1.7     & - & x & + & + & cool   \\
01-12-14 09:06 & 09:06  &  75  & 24h  &-& 2.1 & 14.12:02 & 3.1-1.5     & - & + & + & + & coronal       \\
01-12-28 11:30 & 12:54  &30-40 & 7.5h &?& 2.2 & 28.15:04 & 1.7-1.5     & - & + & + & + &  "  \\

\enddata
\end{deluxetable}
\end{document}